\documentclass[sigconf]{acmart}

\usepackage{multirow}
\usepackage{longtable}
\usepackage{xspace}

\usepackage{enumitem}
\usepackage{changepage} 

  {\list{}{\selectfont\itshape\leftmargin=0.25in\rightmargin=0.25in}\item[]}%
  {\endlist}

\AtBeginDocument{%
  \providecommand\BibTeX{{%
    \normalfont B\kern-0.5em{\scshape i\kern-0.25em b}\kern-0.8em\TeX}}}

\newcommand{\SystemName}{\textsc{BlendScape}\xspace}

\copyrightyear{2024}
\acmYear{2024}
\setcopyright{acmlicensed}\acmConference[UIST '24]{The 37th Annual ACM Symposium on User Interface Software and Technology}{October 13--16, 2024}{Pittsburgh, PA, USA}
\acmBooktitle{The 37th Annual ACM Symposium on User Interface Software and Technology (UIST '24), October 13--16, 2024, Pittsburgh, PA, USA}
\acmDOI{10.1145/3654777.3676326}
\acmISBN{979-8-4007-0628-8/24/10}

\begin{document}

\title[BlendScape: Enabling End-User Customization of Video-Conferencing Environments]{BlendScape: Enabling End-User Customization of Video-Conferencing Environments through Generative AI}

\author{Shwetha Rajaram}
\authornote{This work was done while the first two authors were interns at Microsoft Research. Both authors contributed equally to the paper.}
\affiliation{
  \institution{Microsoft Research}
  \country{United States} 
}
\email{shwethar@umich.edu}

\author{Nels Numan}
\authornotemark[1]
\affiliation{
  \institution{Microsoft Research}
  \country{United States} 
}
\email{nels.numan@ucl.ac.uk}

\author{Balasaravanan Thoravi Kumaravel}
\affiliation{
  \institution{Microsoft Research}
  \country{United States} 
}
\email{bala.kumaravel@microsoft.com}

\author{Nicolai Marquardt}
\affiliation{
  \institution{Microsoft Research}
  \country{United States} 
}
\email{nicmarquardt@microsoft.com}

\author{Andrew D. Wilson}
\affiliation{
  \institution{Microsoft Research}
  \country{United States} 
}
\email{awilson@microsoft.com}

\renewcommand{\shortauthors}{S. Rajaram, N. Numan, B. Kumaravel, N. Marquardt, A. Wilson}

\begin{abstract}
Today’s video-conferencing tools support a rich range of professional and social activities, but their generic meeting environments cannot be dynamically adapted to align with distributed collaborators’ needs. 
To enable end-user customization, we developed \SystemName, a rendering and composition system for video-conferencing participants to tailor environments to their meeting context by leveraging AI image generation techniques. 
\SystemName supports flexible representations of task spaces by blending users’ physical or digital backgrounds into unified environments and implements multimodal interaction techniques to steer the generation. 
Through an exploratory study with 15 end-users, we investigated whether and how they would find value in using generative AI to customize video-conferencing environments.
Participants envisioned using a system like \SystemName to facilitate collaborative activities in the future, but required further controls to mitigate distracting or unrealistic visual elements.
We implemented scenarios to demonstrate \SystemName's expressiveness for supporting environment design strategies from prior work and propose composition techniques to improve the quality of environments.

\end{abstract}



\begin{teaserfigure}
  \vspace{-0.7pc}
  \includegraphics[width=\textwidth]{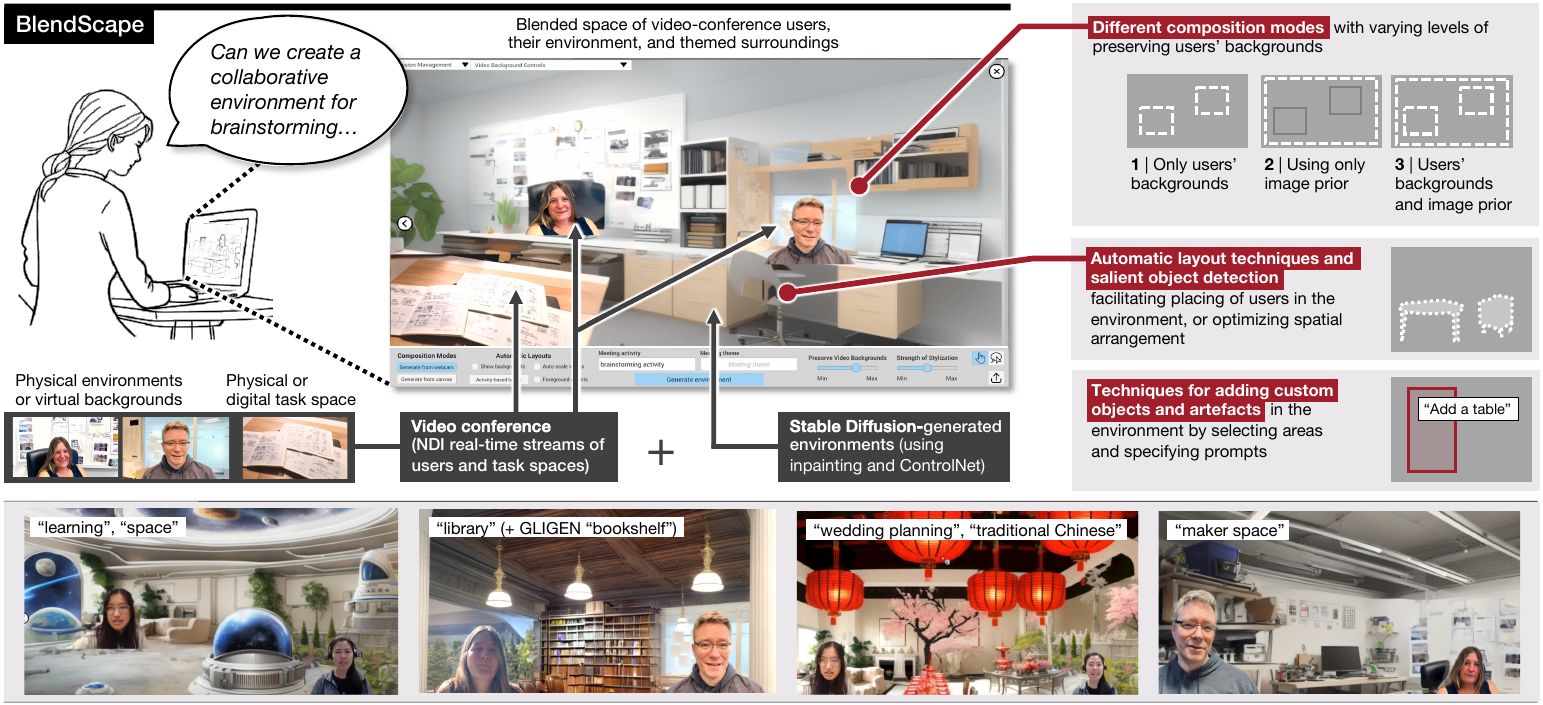}
  \vspace{-1.4pc}
  \caption{Overview of \SystemName, a rendering and composition system for end-users to customize video-conference environments by leveraging AI image generation techniques.}
  \Description{Overview figure of the key components, findings, and visuals results of BlendScape with the following components: (1)  A high-level view of the BlendScape user interface, showcasing a user entering a prompt, a generated environment with composited user video feeds, and a control row for environment adjustments; (2) A summary of key system features: (a) Different composition modes, (b) automatic layout techniques based on object detection, and (c) techniques for adding custom objects and artifacts; (3) A summary of the findings from our evaluation (a) preserving backgrounds enables personal and natural environments, (b) spatial richness vs. cohesiveness of composition, (c) varied preferences for strong vs. subtle theming, (d) context-informed layouts for co-presence, (e) Full generation for social context, subtle composition for professional settings, (f) Potential distractions and required time to generate; (4) A selection of example environments generated by BlendScape.}
  \label{fig:teaser}
\end{teaserfigure}

\maketitle

\section{Introduction}
Advances in video-conferencing technologies and their increasing availability over the past few decades have enabled distributed users to collaborate on activities that previously required face-to-face interaction.
Since the COVID-19 pandemic, video-conferencing has gained popularity not just for facilitating professional tasks (e.g., remote work and distance learning~\cite{Benabdallah-CHI21, Lee-CHI22}), but also for health appointments~\cite{Han-CHI23}, social gatherings~\cite{Fuchsberger-CHI21}, and hobbies~\cite{Cai-CHI21, Yuan-CHI21}. 
However, today's video-conferencing tools do not reflect the rich range of activities that they are used for, due to how they compose \textit{meeting environments} (i.e., the ``stage'' or background rendered around users' videos).
Users are typically placed in separate regions of a video grid within generic meeting rooms, 
which can lead to meeting fatigue~\cite{Fauville-2021-ZoomExhaustion}, reduce user engagement~\cite{Choi-21-AestheticFlattening}, and disrupt interpersonal cues for mediating conversations~\cite{Tang-CSCW23, Hu-CHI23-OpenMic}.

To support more expressive video-conferencing environments that are aligned with distributed collaborators' needs, we envision leveraging generative AI to enable end-users to create custom meeting environments.
To understand the existing design space, we reviewed video-mediated communication research that redesigned meeting spaces to mitigate challenges with distributed collaboration (e.g., communication barriers, decreased sense of co-presence).
We identify three main design strategies:
\textit{(1)}~Establishing the meeting context through the environment (e.g., by rendering shared task spaces~\cite{Gronbaek-CHI21, Hu-CHI23-ThingShare} or thematic visuals~\cite{Hunter-CHI14, follmerVideoPlayPlayful2010});
\textit{(2)}~Leveraging spatial metaphors to enhance communication (e.g., facilitating turn-taking via proxemic interactions between users~\cite{Hu-CHI23-OpenMic});
\textit{(3)}~Using the environment to record a meeting history, to aid future collaboration~\cite{Xia-Spacetime-UIST18}.
Despite the HCI community's knowledge of effective meeting environment designs and empirical studies demonstrating their benefits for distributed collaboration, there is a lack of tool support for end-users to implement these designs in real-time.
Commercial customization tools\footnote{\textbf{Microsoft Teams TogetherMode:} https://www.microsoft.com/en-us/microsoft-teams/teams-together-mode; \textbf{Ohyay:} https://ohyay.co/} require significant manual effort, making it infeasible to adapt environments as meetings progress~\cite{Gronbaek-UIST23}.

\paragraph{\textnormal{As a step towards this vision, we developed \textbf{\SystemName, a rendering and composition system for video-conferencing participants to create environments tailored to their meeting context} (Fig.~\ref{fig:teaser}).}}
We introduce two key innovations: 
\textit{(1)}~We ground the generation of meeting environments in real spaces that are meaningful to users by blending their physical or virtual backgrounds into a unified environment. 
This can serve as a mechanism for personalization~\cite{Venolia-MobileHCI18} or to incorporate physical objects to collaborate around~\cite{Hu-CHI23-ThingShare, Junuzovic-CHI12}.
\textit{(2)}~Capitalizing on recent advances in generative AI, we leverage image generation models to enable expressive and rapid techniques for composing environment designs.
While such techniques are the subject of ongoing research, several dominant modes have emerged:
\begin{itemize}[topsep=6pt]
    \item \textit{text-to-image}: generating an image from a given text prompt.
    \item \textit{image-to-image}: generating an image from a text prompt and an \textit{image prior} (i.e., input image), retaining features of the image prior while introducing new elements or styles consistent with the prompt.
    \item \textit{inpainting}: similar to \textit{image-to-image}, but using a mask to determine which parts of the image prior should be unchanged. The rest of the image is generated in a way that it is consistent with the fixed parts of the image prior (i.e., blended).
\end{itemize}

\paragraph{\textnormal{\SystemName uses \textit{inpainting} to merge users' video backgrounds into blended environments and \textit{image-to-image} techniques to transform existing images of environments to reflect the meeting purpose.}}
To lower the barrier for end-users to generate good quality scenes, we developed multimodal interaction techniques to steer the generation of relevant visuals and composition techniques to naturally integrate users' videos within the scene.

We assessed the benefits and limitations of \SystemName's customization techniques in two steps.
First, to demonstrate the expressiveness, we implemented three scenarios using \SystemName, exploring a range of professional and social collaborative activities.
These scenarios incorporate a majority of environment design strategies for supporting distributed collaboration from our review of prior video-conferencing systems. 

Second, we conducted an exploratory study with 15 end-users to investigate whether and how they would find value in using generative AI to customize video-conferencing environments.
Through guiding participants to prototype meeting spaces for three scenarios using \SystemName, we elicited their customization preferences and explored to what extent \SystemName enabled them to achieve their design intentions.
All participants could envision using generative AI techniques to facilitate a range of collaborative activities in the future (e.g., to spark creativity in professional settings or set a theme for social gatherings).
However, to feel comfortable adapting environments during live meetings, they would require further controls to mitigate distracting or unrealistic visual elements.
We propose improvements to \SystemName's implementation to address these limitations in future work.

Our key contributions are:
\textit{(1)}~the \SystemName composition system, which enables real-time end-user customization of video-conferencing environments through generative AI-driven composition techniques;
\textit{(2)}~an evaluation of \SystemName's expressiveness and considerations for empowering new design participants~\cite{Olsen-UIST07}, through a study demonstrating how 15 video-conferencing users envision leveraging generative AI to personalize meeting environments and our implementation of three target use cases.

\section{Related Work}
Our work extends prior research on 
\textit{(1)}~designing video-conferencing environments to support distributed collaboration;
\textit{(2)}~generative AI techniques for constructing 2D and 3D environments.

\subsection{Design Strategies for Video-Conferencing Environments}
\label{sec:rel-work-vidconf-envs}
Our goal with \SystemName was to provide a unified set of customization techniques for dynamically tailoring meeting environments to a wide range of collaborative activities.
Here, we define the \textit{environment} as the ``stage'' or background rendered around users' videos, excluding functional tools such as the chat or host controls.

\begin{figure}[htb!]
    \centering
    \includegraphics[width=\linewidth]{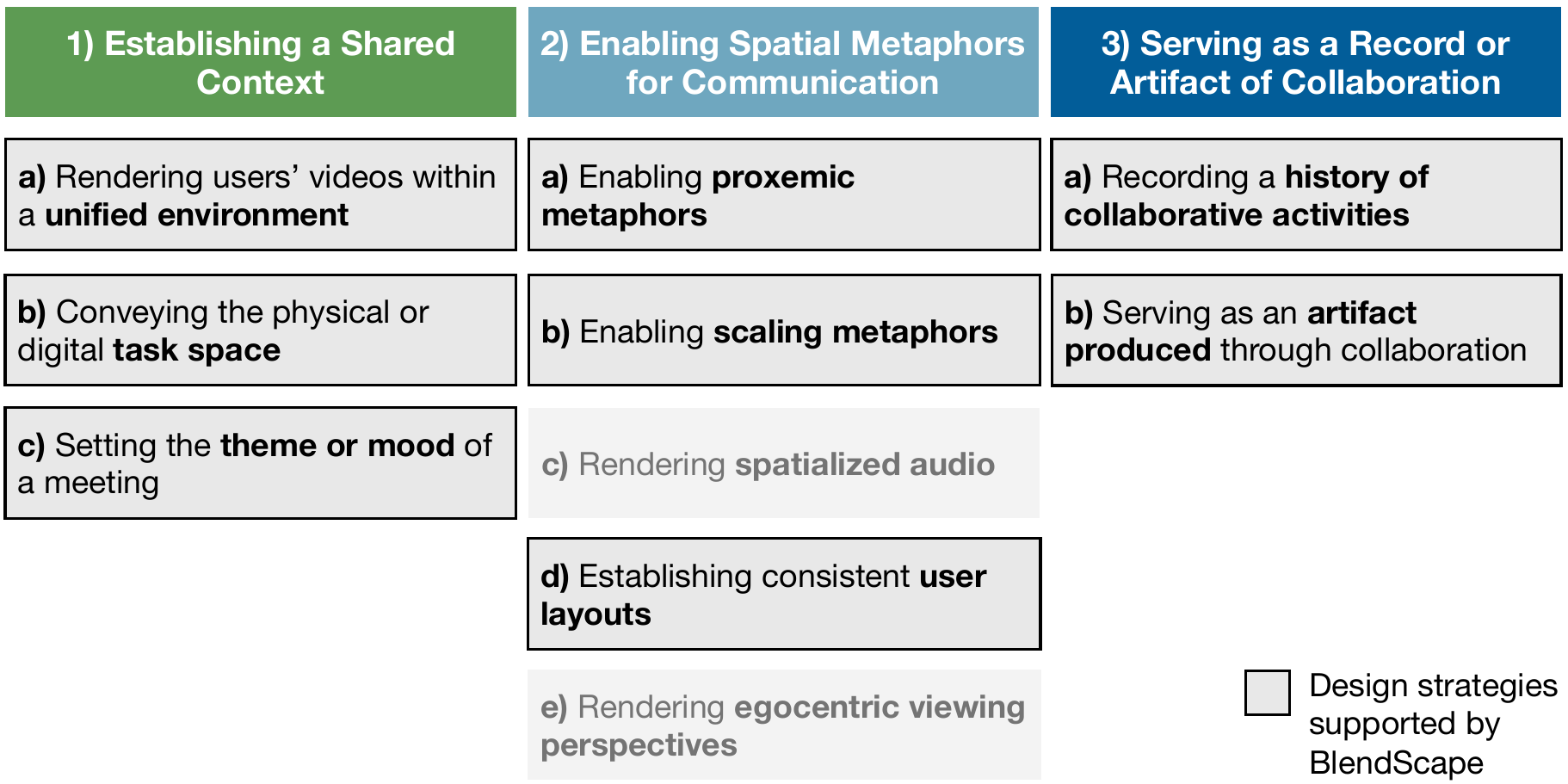}
    \caption{Classification of Environment Design Strategies: We analyzed how existing video-conferencing tools compose meeting spaces to support distributed collaboration by \textit{(A)} depicting a shared context, \textit{(B)} enhancing communication behaviors through spatial metaphors, \textit{(C)} capturing a record of collaboration within the space.
    In Sec.~\ref{sec:scenario-demonstration}, we use scenarios to demonstrate how \SystemName supports implementing eight of these ten design strategies (shown in bold).}
    \Description{Classification of Environment Composition Techniques. This figure shows a table with three columns. The first column has header "Establishing a Shared Context" and children "Rendering users' videos within a unified environment", "Conveying the physical or digital task space", "Setting the theme or mood of a meeting. The second column as header "Enabling Spatial Metaphors for Communication" and children "Enabling proxemic metaphors", "Enabling scaling metaphors", "Rendering spatialized audio", "Establishing consistent user layouts", "Rendering egocentric viewing perspectives." The third column has header "Serving as a Record or Artifact of Collaboration" and children "Recording a history of collaborative activities", "Serving as an artifact produced through collaboration."}
    \label{fig:rel-work-env-composition}
\end{figure}

To understand the design space of meeting environments, 
we reviewed commercial video-conferencing tools and systems from video-mediated communication research. 
We classified the environment design strategies these systems adopted to mitigate challenges in distributed collaboration, as demonstrated through empirical studies. 
Our review surfaced three main roles that meeting environments can play in supporting distributed collaboration (Fig.~\ref{fig:rel-work-env-composition}):
\textit{(1)}~establishing a shared context, \textit{(2)}~enabling spatial metaphors for communication, \textit{(3)}~serving as a record or artifact of collaboration.

We focus our review on screen-based meeting systems with 2D or 2.5D designs. 
Design strategies for other modalities (e.g., tabletop~\cite{Zagermann-CHI16, Tuddenham-CHI09} or mixed reality interfaces~\cite{Pejsa-CSCW16, Herskovitz-ISS22, Gronbaek-CHI23}) would likely introduce new dimensions to our categorization.

\paragraph{\textnormal{\textbf{Establishing a Shared Context.}}}
A primary role of meeting environments is creating a shared frame of reference for distributed users.
We observed three design strategies (Fig.~\ref{fig:rel-work-env-composition}A): displaying users within a unified meeting space, incorporating elements of physical or digital task spaces, and establishing a meeting theme.

Simulating co-located meetings by \textbf{rendering users' videos within a unified environment} is a popular strategy to increase distributed users' sense of co-presence~\cite{Tang-CSCW23, Hunter-CHI14}, as demonstrated both by commercial tools (e.g., Teams Together Mode, Ohyay\footnotemark[1]) and research systems (e.g., Waazam~\cite{Hunter-CHI14}, HyperMirror~\cite{Morikawa-CHI98}, BISi~\cite{OHara-TOCHI11}).
\textbf{Conveying the task space}, i.e., the physical or digital space where artifacts are collaboratively produced~\cite{Buxton2009}, is critical for distributed users who would otherwise lack awareness of their collaborators' actions.
Beyond traditional screen-sharing capabilities that display \textit{digital task spaces},
MirrorBlender's~\cite{Gronbaek-CHI21} layerable ``mirrors'' (translucent representations of users' videos and shared screens) enable more natural interactions with digital artifacts, e.g., using hands to gesture around shared content.
Capturing \textit{physical task spaces} (e.g., for learning hands-on skills~\cite{Kumaravel-UIST19}) often requires custom hardware setups~\cite{Junuzovic-CHI12, Labrie-GROUP22}.
To mitigate this, ThingShare~\cite{Hu-CHI23-ThingShare} enables users to scan physical objects via webcams and manipulate their digital representations. 
To \textbf{set a common theme or mood for meeting participants}, Wazaam~\cite{Hunter-CHI14} and VideoPlay~\cite{follmerVideoPlayPlayful2010} introduce capture and rendering techniques for playful interactions, e.g., compositing children within storybook illustrations.

\paragraph{\textnormal{\textbf{Enabling Spatial Metaphors for Communication.}}}
To enable more natural and seamless communication, recent tools leverage spatial affordances to mimic collaboration strategies from face-to-face interactions.
Our review surfaced five spatial composition techniques (Fig.~\ref{fig:rel-work-env-composition}B): enabling conversational transitions through manipulating proximity or scale of users' videos, rendering spatial audio, structuring users' layouts to support turn-taking, and rendering egocentric viewing perspectives.

First, meeting rooms that resemble physical spaces afford using \textbf{proxemic interactions} to facilitate conversations: in Gather\footnote{\textbf{Gather:} https://www.gather.town/}, users can initiate one-on-one video calls by ``walking up'' to each other; in OpenMic~\cite{Hu-CHI23-OpenMic}, users position themselves near a ``Virtual Floor'' to express their desire to speak.
\textbf{Scaling metaphors} are commonly used to highlight specific users: Teams' and Zoom's Speaker views render active speakers at a larger size than other meeting participants; OpenMic~\cite{Hu-CHI23-OpenMic} allows users to negotiate conversational transitions by increasing the size of their videos.
\textbf{Rendering spatial audio} can help users follow conversation flows and support inclusion of remote participants in hybrid meetings~\cite{Hyrkas-CHI23, Mu-WhisperThroughWalls-CHI24} 
(e.g., MirrorVerse's~\cite{Gronbaek-UIST23} ``doorway'' function gives users an auditory preview of breakout room discussions before they join).
\textbf{Maintaining consistent user layouts}, (e.g., by seating users around a table in TogetherMode\footnotemark[1]) can help establish turn-taking patterns.
Perspectives~\cite{Tang-CSCW23} further supports turn-taking by \textbf{rendering egocentric viewpoints} in the environment for each user, which simulates face-to-face social cues, e.g., making eye contact with speakers.

\paragraph{\textnormal{\textbf{Serving as a Record or Artifact of Collaboration.}}}
Finally, we observed two examples where meeting environments document collaborative activities, providing a basis for later ideation (Fig.~\ref{fig:rel-work-env-composition}C). 
First, meeting tools can \textbf{record a history of user interactions, movements, and changes to the meeting environments}, which can be replayed at a later time to understand collaboration patterns (e.g., MirrorVerse's workspace record-and-replay tools~\cite{Gronbaek-UIST23}). 
Finally, the meeting environment could \textbf{be an artifact produced through users' collaboration}, e.g., for interior design or world-building scenarios. 
In our review of prior work, we did not find examples of video-conferencing tools that explicitly claim this functionality; however, this use case is common in VR immersive authoring tools~\cite{Xia-Spacetime-UIST18, Zhang-VRGit-CHI23} which allow users to collaboratively create virtual content while situated in the virtual world itself. 

\paragraph{\textnormal{While the HCI community has significant empirical knowledge on how to configure meeting environments to enhance distributed collaboration, it is still a challenge for end-users to attain these designs, due to a lack of real-time customization support in video-conferencing tools~\cite{Gronbaek-UIST23}.}}
Commercial authoring tools (e.g., from Together Mode\footnotemark[1], Ohyay\footnotemark[1], and Gather\footnotemark[2]) are geared towards pre-meeting use, requiring users to manually craft relevant visuals and establish user layouts.
Inspired by recent customization suites like MirrorVerse~\cite{Gronbaek-UIST23}, our work aims to lower the barrier for users to dynamically create video-conferencing environments.
We leverage generative AI techniques as a new approach for rapidly expressing and aligning environment designs with meeting contexts (e.g., to visualize shared task spaces or create themed visuals).

\subsection{AI-Assisted Environment Generation}
Our work also builds on prior approaches for enabling non-technical users to interact with generative AI models, which has been the focus of recent HCI research~\cite{agarwalaInteractiveDigitalPhotomontage2004, el-noubyTellDrawRepeat2019, dangGANSliderHowUsers2022, liuOpal2022,dangWorldSmithIterativeExpressive2023}. 
Based on user input, such as visuals or text-based prompts, these models can generate a wide variety of content including images~\cite{avidanSeamCarvingContentaware2007,mirzaConditionalGenerativeAdversarial2014,zhangRealtimeUserguidedImage2017,yuFreeFormImageInpainting2019,rombachHighResolutionImageSynthesis2022}, text~\cite{radfordImprovingLanguageUnderstanding2018,brownLanguageModelsAre2020,openaiGPT4TechnicalReport2023}, and even 3D objects~\cite{coyneWordsEyeAutomaticTexttoscene2001,gkioxariMeshRCNN2020,kohSimpleEffectiveSynthesis2022,karnewarHOLODIFFUSIONTraining3D2023,junShapEGeneratingConditional2023}.

Recent advances in image generation techniques allow the generation to be conditioned in a variety of useful ways, in addition to guidance from text-based prompts. 
For example, ControlNet uses an auxiliary model that incorporates additional data such as depth, semantic, and human pose representations~\cite{zhangAddingConditionalControl2023,liGligenOpensetGrounded2023}. 
Prior work also explored intuitive techniques for end-users to control generative AI models, e.g., through
sketching~\cite{chungTaleBrushSketchingStories2022,zhangAddingConditionalControl2023, chungPromptPaintSteeringTexttoImage2023}, speech~\cite{numanUbiqGenieLeveragingExternal2023}, sliders~\cite{chungArtinterAIpoweredBoundary2023}, and iterative design mechanisms~\cite{sharmaChatPainterImprovingText2018,el-noubyTellDrawRepeat2019,dangWorldSmithIterativeExpressive2023}.

There is increased interest in harnessing image generation models to dynamically generate virtual environments. 
Before the rise of AI-driven approaches, environment generation relied on predefined components and procedural logic. 
For example, \textit{WordsEye} converts text descriptions into 3D scenes using a database of 3D models and predefined rules~\cite{coyneWordsEyeAutomaticTexttoscene2001}. 
\textit{DreamWalker} substitutes real-world elements with virtual content via procedural generation, enabling VR users to safely navigate their physical surroundings~\cite{yangDreamWalkerSubstitutingRealWorld2019}.

These procedural approaches paved the way for the utilization of generative techniques that can be observed in recent AI-assisted creativity support tools.
A recent example is \textit{WorldSmith}, a system that investigates how multimodal image generation models can be harnessed to aid users in authoring and iteratively refining elements of fictional worlds~\cite{dangWorldSmithIterativeExpressive2023}. \textit{Opal}~\cite{liuOpal2022} guides users through a structured search for visual concepts to generate images for news illustrations, utilizing LLMs to tune users' prompts based on an article's content.

\SystemName builds on similar techniques, but with a distinct focus on generating creative environments for video-conferencing systems that can be aligned with meeting participants' goals. 
In particular, we employ \textit{inpainting} methods to blend users' video backgrounds into unified scenes, provide multimodal input techniques for users to steer the generation, and leverage LLMs to dynamically tailor users' prompts to meeting themes and activities.

\begin{figure}[htb!]
    \centering
    \includegraphics[width=\linewidth]{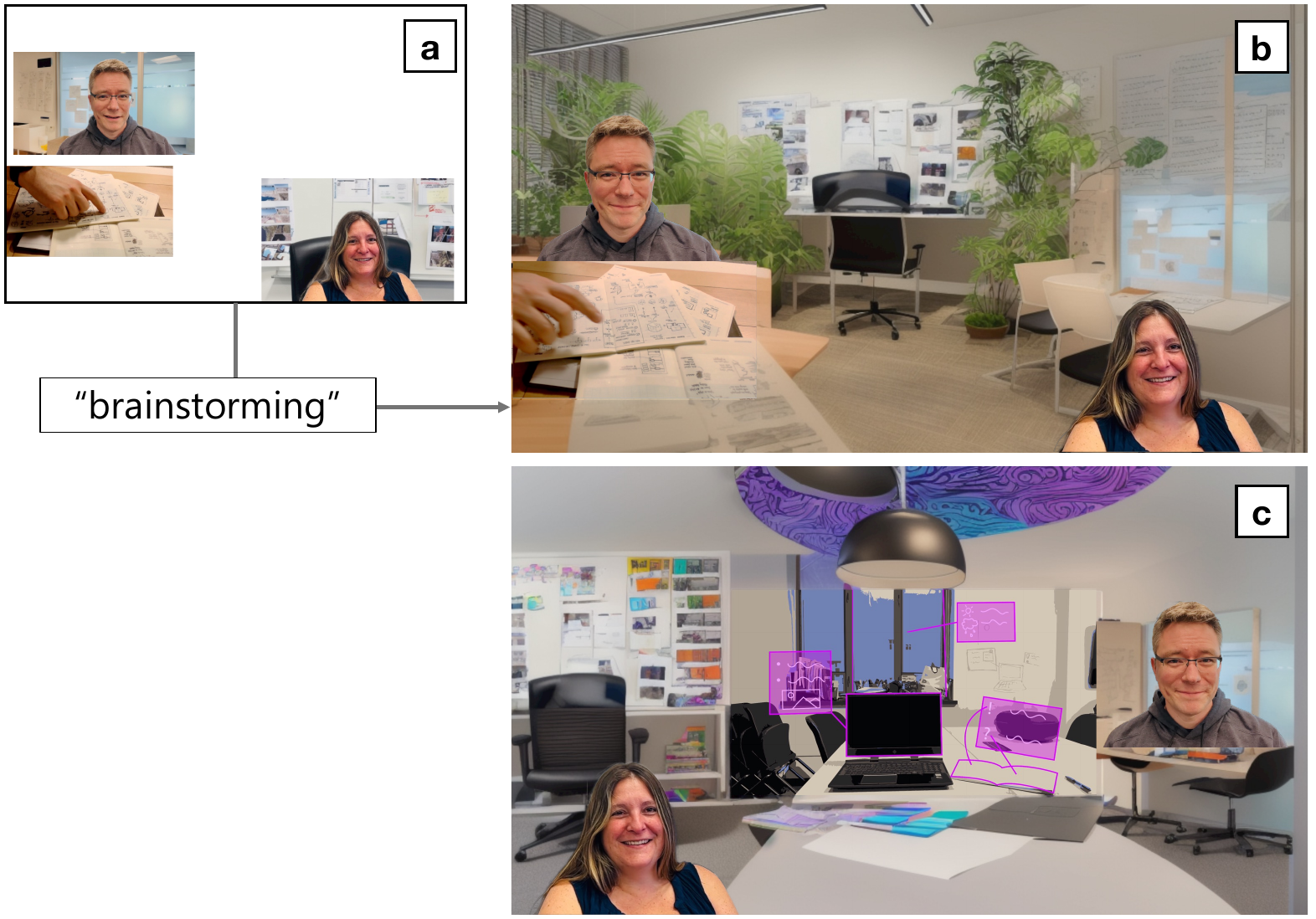}
    \caption{Scenario 1: \textit{Design Brainstorming}. To create a unified setting for brainstorming, two designers use \SystemName to blend their webcam backgrounds with a camera feed of a physical desk \textit{(a, b)}, enabling them to ideate around hand-drawn sketches. They later blend in elements of their digital task space, such as mock-ups of a mixed reality interface \textit{(c)}.}
    \Description{Overview of "Product Design Ideation" scenario with the following components: (a) Blank canvas with original user video feeds. (b) Blended environment with user video feeds composited into the scene based on prompt "brainstorming", with region-based user-defined prompt specifying "wall with post-it notes" for granular environment editing. (c) Blended environment with addition of wall with post-it notes according to user input. (d) Blended environment with user video feeds composited into the scene based on prompt "design studio with the design of sketches of prototypes".}
    \label{fig:scenario1}
    \vspace{-1em}
\end{figure}

\section{Scenario Walkthrough}
\label{sec:initial-scenario}
To illustrate how \SystemName can enable expressive meeting environments by blending physical and digital spaces, we present a scenario implemented with our system: a brainstorming session between two designers who are prototyping a mixed reality interface.

The designers join a video call and add their webcam feeds to \SystemName's composition interface. 
One designer also adds a camera feed of their physical notebook, so they can sketch ideas during the brainstorming session (Fig.~\ref{fig:scenario1}a). 
At the start of the meeting, they use \SystemName to generate a creative environment for their ideation activity. 
They type ``\texttt{brainstorming}'' in the prompt field, and
a few seconds later, they see their \textit{blended environment}: their physical surroundings are still visible, but now seamlessly extend into a unified design studio (Fig.~\ref{fig:scenario1}b).
This blended meeting space allows them to gesture and refer to physical sketches in the notebook, simulating how they might collaborate face-to-face.
Later in the meeting, the designers blend a digital sketch into their physical backgrounds to preview mock-ups of their MR interface (Fig.~\ref{fig:scenario1}c).

Many other meeting experiences are possible, from work scenarios to therapy spaces, birthday parties, or vacation planning (Appendix.~\ref{appendix:GPT-aided-prompting}). 
As we describe next, \SystemName provides flexible techniques for creating rich meeting spaces.
Later, we present additional scenarios that demonstrate \SystemName's expressiveness to enable distributed collaboration techniques from prior work ~(Sec.~\ref{sec:scenario-demonstration}).

\begin{figure*}[ht!]
    \centering
    \includegraphics[width=0.75\linewidth]{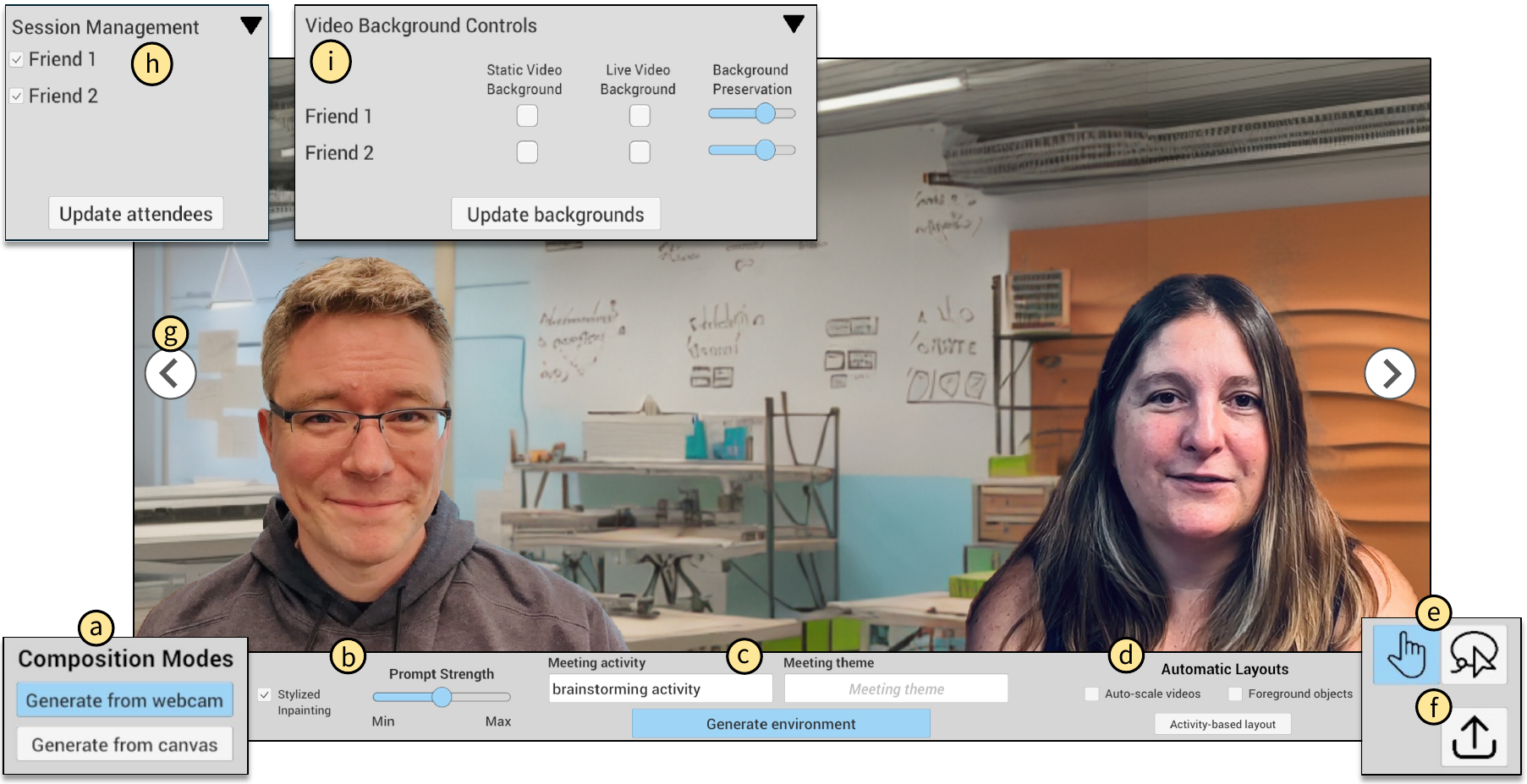}
    \caption{Overview of \SystemName interface: \SystemName offers two composition modes for creating meeting spaces \textit{(a)}:~blending webcam feeds together via \textit{inpainting} and transforming the image on the canvas via \textit{image-to-image}.
    To steer the environment generation, end-users can specify text-based prompts for the \textit{Meeting Activity} and \textit{Meeting Theme}~\textit{(c)}, control the strength of stylistic prompts~\textit{(b)}, upload custom image priors~\textit{(f)}, and modify specific regions of the scene via selection tools~\textit{(e)}.
    Users can return to and iterate on previous environment designs via the history tools~\textit{(g)}.
    The automatic layout techniques facilitate positioning users behind foreground objects in the scene~\textit{(d)}. 
    \SystemName also provides session management tools~\textit{(h)} and per-user controls for adjusting the proportion of their video backgrounds preserved during the environment generation and toggling between displaying live webcam feeds or static frames \textit{(i)}.
    }
    \Description{Overview of BlendScape's user interface: (a) Two composition modes – inpainting using a webcam feed and image-to-image techniques for refining canvas images, (b) Automatic layout techniques for compositing users behind objects and scaling videos, (c) Input fields for prompts specifying Meeting Activity and Meeting Theme, (d) Slider for users to specify the extent of video background preservation, (e) Strength of Stylization slider to control the priority of stylistic prompts or user backgrounds, (f) Transform tool for moving and scaling user video feeds, (g) Selection tool to mark area for region-based modification, and (h) Button to upload an image to the canvas.}
    \label{fig:system-walkthrough}
\end{figure*}

\section{BlendScape System}

This section presents \SystemName, a rendering and composition system that enables meeting participants to customize video-conferencing environments. 
Key to our approach is blending elements of users' physical or virtual backgrounds into unified environments to allow for flexible representations of task spaces.
We first outline three system requirements that guided the design of our system.   
Then, we describe our implementation of \SystemName's generative AI techniques for customizing meeting environments and composition techniques for enhancing visual cohesion.

\subsection{Requirements}
We defined three requirements for customizing meeting environments with \SystemName, informed by our review of environment composition techniques for distributed collaboration (Sec.~\ref{sec:rel-work-vidconf-envs}) and our motivating scenarios (Sec.~\ref{sec:initial-scenario}, ~\ref{sec:scenario-demonstration}):

\textbf{R1: Enabling users to express the meeting context through the environment.}
Embedding relevant visual and structural elements within video-conferencing environments can strengthen distributed users' collaborative processes (e.g., conveying shared task spaces in our \textit{Design Brainstorming} scenario, using spatial landmarks to facilitate conversational transitions~\cite{Hu-CHI23-OpenMic, Tang-CSCW23}).
\SystemName allows users to align environments to their meeting purpose by specifying text-based prompts and providing \textit{image priors} (i.e., images of their physical surroundings or other spaces that represent their collaboration needs). 

\textbf{R2: Supporting convincing illusions of meeting in a shared space.}
To simulate face-to-face communication cues (e.g., deictic gestures~\cite{Gronbaek-CHI21}, making eye-contact~\cite{Tang-CSCW23}), prior systems composite users within virtual environments that resemble physical spaces. 
To enable these designs while allowing users to incorporate their physical context, \SystemName implements two rendering techniques: \textit{(1)}~preserving and blending users' video backgrounds into a unified environment, which maintains realistic lighting, color temperature, and shadows around users; \textit{(2)}~\textit{hidden surface removal} to obscure the boundaries of users' webcams among objects in the scene.

\textbf{R3: Enabling coarse- and fine-grained customization of environments.}
Adapting meeting spaces to dynamic collaboration needs may require users to update the entire scene (e.g., to reconfigure the space for small vs. large group discussions~\cite{Gronbaek-UIST23}), as well as make minor adjustments (e.g., incorporating shared content~\cite{Hu-CHI23-ThingShare}).
In addition to \textit{inpainting} and \textit{image-to-image} techniques for creating new environments, \SystemName enables users to make granular changes by selecting portions of the scene to re-generate.
To enable iteration on past results, we maintain a history of environments.

\begin{figure*}[h!]
    \centering
    \includegraphics[width=0.9\linewidth]{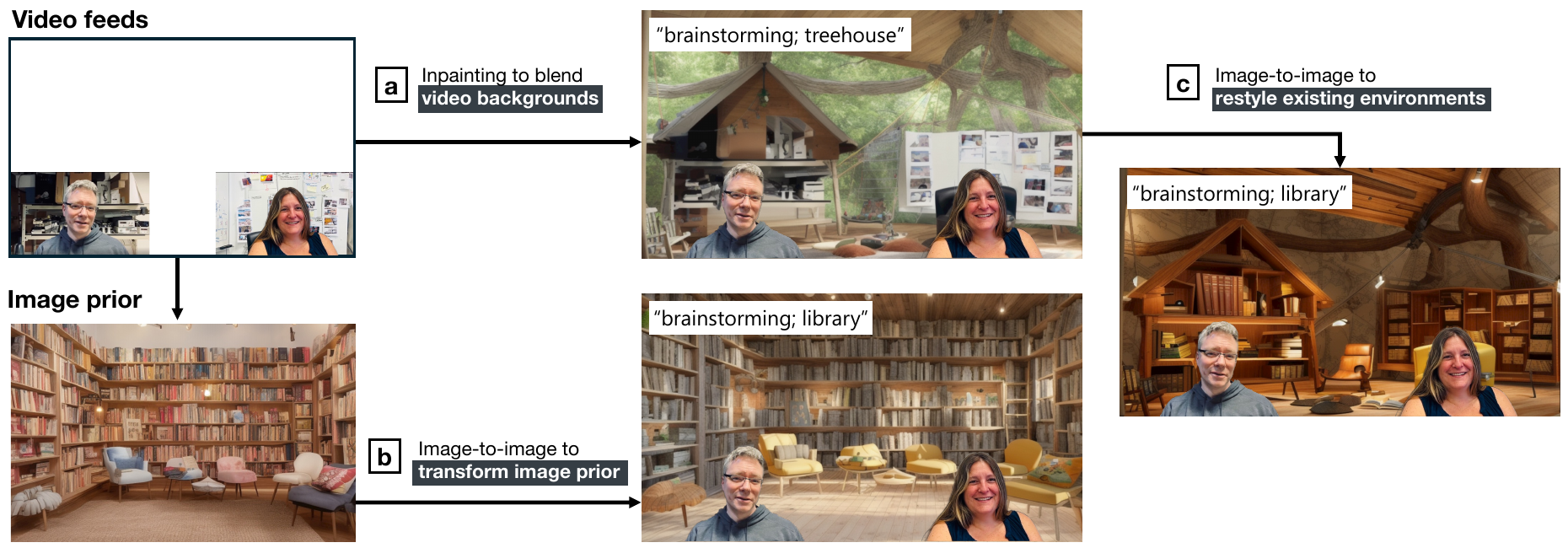}
    \caption{Environment Generation Techniques.
    \SystemName supports composing meeting spaces through \textit{(a)} blending video feeds together via \textit{inpainting} techniques and \textit{(b)} transforming an input image (i.e., image prior) via \textit{image-to-image} techniques. These composition approaches can be chained, e.g., to restyle a blended environment in the theme of a library \textit{(c)}.
    }
    \label{fig:image-generation-approaches}
    \Description{An overview of environment generation techniques in a flow diagram. BlendScape supports composing meeting spaces through (a) blending video feeds together via inpainting techniques and (b) transforming an input image (i.e., image prior) via image-to-image techniques. These composition approaches can be chained, e.g., to restyle a blended environment in the theme of a library (c).}
\end{figure*}

\subsection{\SystemName Interface}

Next, we provide an overview of \SystemName's composition tools (Fig.~\ref{fig:system-walkthrough}). 
\SystemName's user interface consists of a canvas in the center that displays blended environments and composites the users' videos within them. 
All meeting participants share the same view of this canvas; changes to the environment and position of users' video feeds are synchronized across all users.

Meeting participants can generate environments using \textbf{two composition modes} (Fig.~\ref{fig:system-walkthrough}a): blending the video feeds from only their webcams (using \textit{inpainting}), or refining the image that is already present in the canvas (using \textit{image-to-image} techniques).
This image on the canvas may be a result of a previous image generation step or an image uploaded by the user (Fig. \ref{fig:system-walkthrough}f).
For \textit{inpainting}, users can specify how much of their video backgrounds should be preserved through a slider (Fig. \ref{fig:system-walkthrough}i).

To generate visuals relevant to the purpose of their meeting, users can enter two \textbf{prompts for a \textit{Meeting Activity} and \textit{Meeting Theme}} (Fig. \ref{fig:system-walkthrough}c).
The \textit{Prompt Strength} slider controls the prompt weight, i.e., how much to prioritize the prompts in the environment generation (Fig. \ref{fig:system-walkthrough}b).
\SystemName offers direct manipulation techniques (clicking, dragging, and pinching) as intuitive ways to position and scale videos, both for steering the environment generation and placing users' videos in the scene.

To make fine-grained adjustments to the environment, users can add or remove objects by selecting a region of the scene (Fig.~\ref{fig:system-walkthrough}e) and specifying a prompt. \SystemName also offers \textbf{automatic layout techniques} (Fig.~\ref{fig:system-walkthrough}d) that composite users behind objects in the scene and automatically scale their videos to match the size of the objects.
\SystemName saves a \textbf{history of past environment generations and users' positions within the scene}, to enable iterating on past designs (Fig.~\ref{fig:system-walkthrough}g).

\subsection{Generative AI Techniques for Blended Environments}
To enable real-time end-user customization of video-conferencing environments, 
\SystemName implements two classes of AI image generation techniques: \textit{(1)} an \textbf{inpainting technique}, which blends users' physical or virtual backgrounds into a unified environment, and \textit{(2)} an \textbf{image-to-image technique}, which incorporates users into an overarching image that represents the meeting setting.

\begin{figure}[h!]
    \centering
    \includegraphics[width=\linewidth]{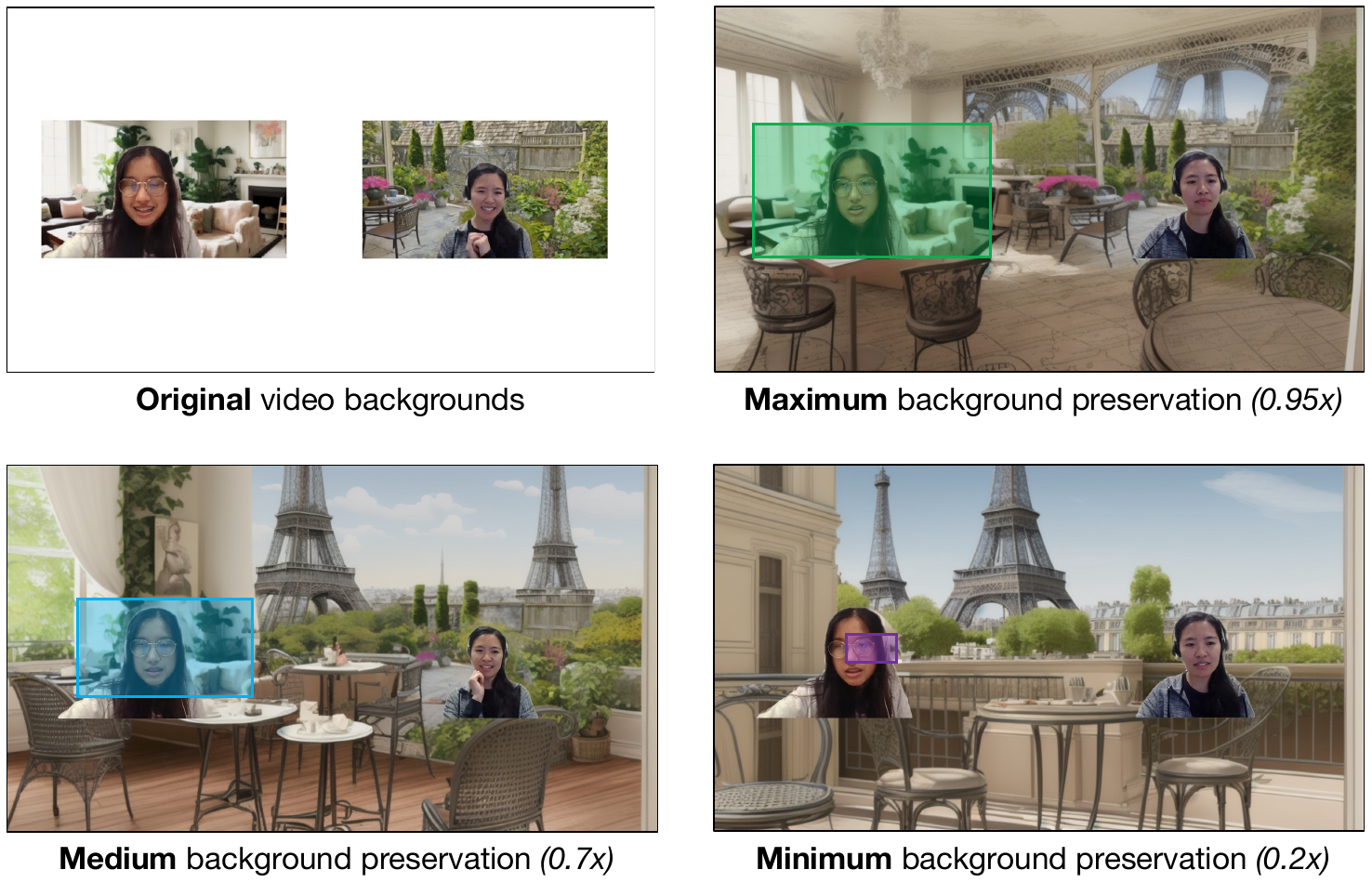}
    \caption{Masking Video Backgrounds: Users can adjust the proportion of their physical or virtual surroundings to retain in the resulting blended environments.
    }
    \label{fig:4-masking}
    \Description{Overview of techniques for masking video backgrounds. Via BlendScape’s video background controls (Fig. 4), users can adjust the proportion of their physical or virtual surroundings to mask or retain in the resulting blended environments. Four images are shown in a 2-by-2 grid. In the top-left image, the original video backgrounds are shown on a blank canvas. In the other three corners, different inpainted canvases are shown. These canvases are generated with different levels of background preservation (minimum, medium, maximum), where minimum results in the least amount of influence of the user’s background video on the generated canvas and maximum refers to the most influence on the generated canvas.}
\end{figure}

\paragraph{\textnormal{\textbf{Generating blended environments from users' video backgrounds}} \textit{(R1, R2)}}
\SystemName allows meeting participants to incorporate their real-world surroundings or virtual backgrounds into the shared environment as a mechanism for personalization~\cite{Venolia-MobileHCI18} or capturing the task space where artifacts are collaboratively produced~\cite{Buxton2009}.
First, users can specify the proportion of their video backgrounds to retain; we mask these regions to preserve them in the blended environment (Fig.~\ref{fig:4-masking}).
Then, \SystemName performs \textit{inpainting} to generate plausible visual details between the fixed regions, based on user-specified prompts for the \textit{Meeting Activity} and \textit{Meeting Theme}, e.g., \texttt{``brainstorming''} in a \texttt{``treehouse''}-themed environment (Fig.~\ref{fig:image-generation-approaches}a). 
By preserving physical backgrounds, we aimed to more naturally integrate users into the blended space by matching their real-world conditions (e.g., lighting, shadows, color temperature, and webcam resolution).

Direct manipulation of video feeds (i.e., re-positioning and re-scaling) can be used to steer the generation of different styles of \textit{inpainted} environments. 
For example, positioning smaller videos at the top of the screen creates a sense of environmental depth, with clear separation of foreground and background elements.

\paragraph{\textnormal{\textbf{Driving environment generation through \textit{image priors}}} \textit{(R1)}}
As a second composition approach, \SystemName uses \textit{image-to-image} techniques to transform an \textit{image prior} (i.e., input image).
This involves restyling the environment to incorporate visuals related to the \textit{Meeting Activity} and \textit{Meeting Theme} prompts, while preserving key structural elements in the prior, e.g., walls or background objects that convey the environment's geometry (Fig.~\ref{fig:image-generation-approaches}b).

Meeting participants can upload image priors to set the theme of the meeting, (e.g., using an image of a library for a study group), establish a layout of users (e.g., row-based seating for academic lectures), or situate themselves in real-world meeting spaces.

\paragraph{\textnormal{\textbf{Combining \textit{inpainting} and \textit{image-to-image} techniques for iterative environment design.}} \textit{(R1, R3)}}
To simplify novice users' composition processes into two clear pathways for creating environments, we separated the \textit{inpainting} and \textit{image-to-image} techniques into two modes within \SystemName: webcam-based and canvas-based generation. 
However, \SystemName supports flexibly combining these techniques to iterate on prior environments.
For example, performing an \textit{image-to-image} transformation after an \textit{inpainting} generation is an effective strategy for refining roughly-blended areas and subtly restyling an existing environment.

\subsection{Improving the Composition of Blended Environments}
\label{sec:composition-techniques}
To make the base environments generated via \textit{inpainting} and \textit{image-to-image} techniques more suitable for video-conferencing, we implemented three composition techniques:
\textit{(1)}~\textbf{prompt enhancement} to further align environments to the meeting context and enhance the visual quality; 
\textit{(2)}~\textbf{granular scene-editing controls} to add visuals or correct distortions;
\textit{(3)}~\textbf{hidden surface removal} techniques to integrate users' videos with foreground objects.

\paragraph{\textnormal{\textbf{Generating context-informed and thematic environments through LLM-driven prompt padding}} \textit{(R1, R2)}}
Crafting expressive prompts remains a challenge for novice users of image generation models, requiring significant trial and error~\cite{bradePromptifyTexttoImageGeneration2023, chungPromptPaintSteeringTexttoImage2023}.
To lower the barrier for meeting participants to create detailed environments, \SystemName enhances user-specified prompts with keywords suggested by LLMs, as demonstrated by prior systems~\cite{liuOpal2022, bradePromptifyTexttoImageGeneration2023}.

First, we elicit text-based prompts from users to define the \textit{Meeting Activity}, which drives the layout of the environment, and the \textit{Meeting Theme}, which steers the generation of aesthetic elements.
Then, to dynamically tailor the image generation prompts to the meeting context, we query GPT-3.5 to augment the \textit{Meeting Activity} and \textit{Meeting Theme} prompts with keywords for five relevant objects and five stylistic qualities that represent the meeting atmosphere. 
For example, for a \textit{Meeting Activity} of ``\texttt{Brainstorming Session}'' and \textit{Meeting Theme} of ``\texttt{Hologram},'' GPT-3.5 suggested ``\texttt{Interactive Touchscreens}'' and ``\texttt{Holographic Whiteboards}'' as objects and ``\texttt{Dynamic Lighting}'' and ``\texttt{Seamless Integration of Virtual \& Physical Elements}'' as stylistic qualities.

Inspired by prompt expansion tools for artists (e.g., PromptGen\footnote{\textbf{PromptGen:} https://promptgen.vercel.app/}), \SystemName also adds a fixed set of terms to encourage high-quality visuals: ``\texttt{highly detailed, intricate, sharp focus, smooth}.''
Appendix~\ref{appendix:GPT-aided-prompting} includes examples of GPT queries and outputs.

\paragraph{\textnormal{\textbf{Compositing users in an immersive manner through \textit{hidden surface removal}}} \textit{(R2)}}
A limitation of capturing meeting participants via webcams is that they appear as ``floating heads'' with a harsh cut-off at their shoulders, due to the limited field-of-view. 
To more naturally integrate users within environments, \SystemName takes a similar approach as prior tools (e.g., Ohyay, TogetherMode\footnotemark[1]): placing users behind objects in the scene and enabling \textit{hidden surface removal}, a rendering technique to remove portions of 3D objects that should not be visible from a particular camera perspective. 
We implemented an object segmentation pipeline in ~\SystemName to compute the salient objects in the generated environment (e.g., chairs, tables) and extract them to a foreground layer positioned in front of users' videos (Fig.~\ref{fig:system-layers}). 
This creates the illusion of users sitting behind objects.

\begin{figure}[htb!]
    \centering
    \includegraphics[width=\linewidth]{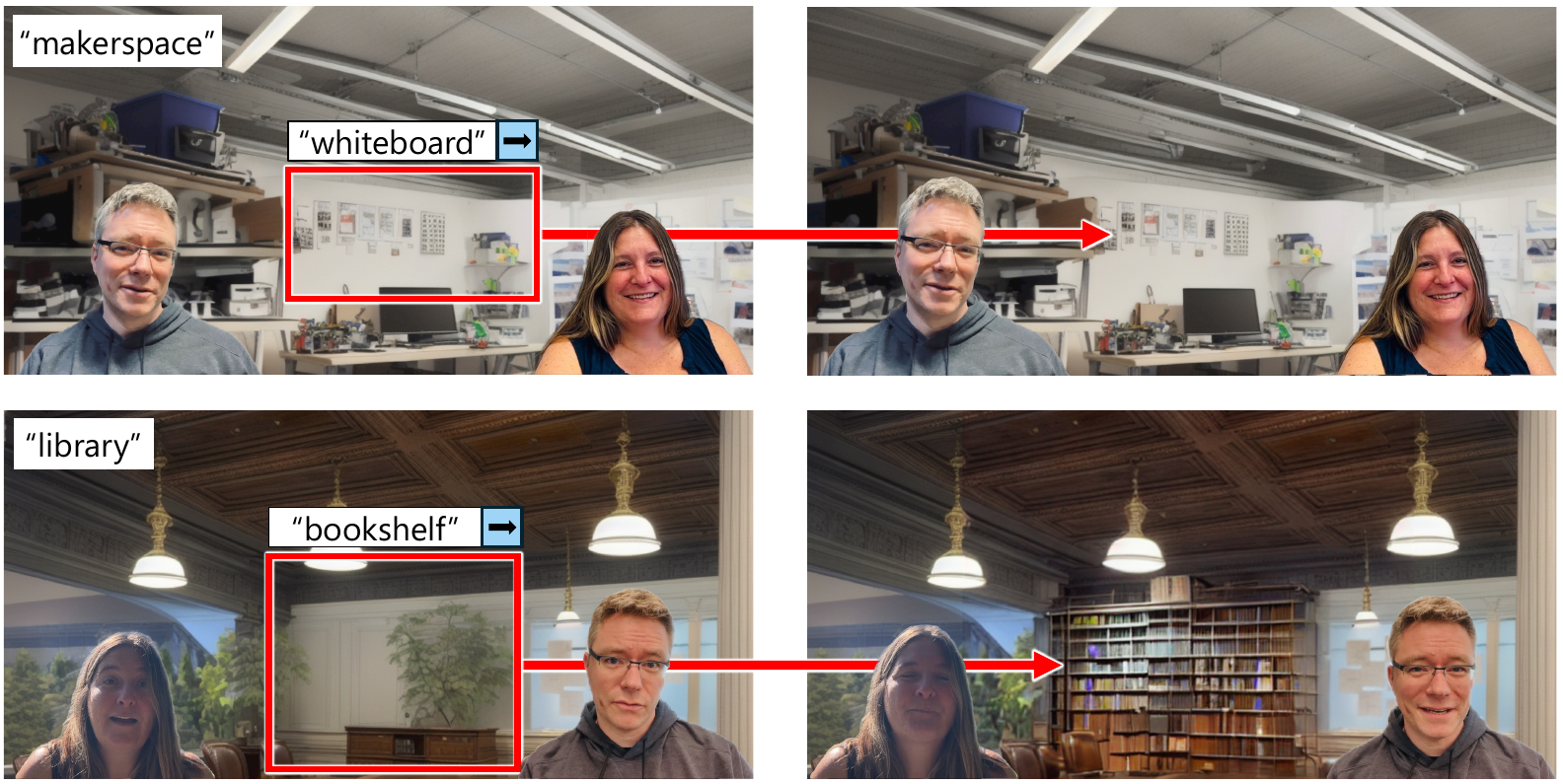}
    \caption{Granular Editing Tools: To add or remove content from the scene, users can outline a region and specify a text-based prompt.}
    
    \label{fig:4-gligen}
    \Description{Region-based modification of blended backgrounds, with two examples. Firstly, a "maker space" blended environment is modified by adding a whiteboard based on a user-defined prompt and region. Secondly, a "library" blended environment is modified by adding a bookshelf based on a user-defined prompt and region.}
\end{figure}

\paragraph{\textnormal{\textbf{Making granular edits 
 to the environment}} \textit{(R3)}}
In addition to \SystemName's \textit{inpainting} and \textit{image-to-image} for updating the entire video-conferencing environment, we implemented a technique for making fine-grained revisions to the scene (Fig.~\ref{fig:4-gligen}). 
Using the selection tool, users can circle areas of the scene and specify a text-based prompt to remove content (e.g., to fix distorted areas) or add visuals relevant to the meeting scenario (e.g., chairs to accommodate new meeting participants). 
To regenerate the area, \SystemName uses the GLIGEN~\cite{liGligenOpensetGrounded2023} inpainting model, which is trained to generate cohesive results by considering the prompt, position, and scale of the specified region in relation to its surroundings.

\begin{figure*}[ht!]
    \centering
    \includegraphics[width=0.9\linewidth]{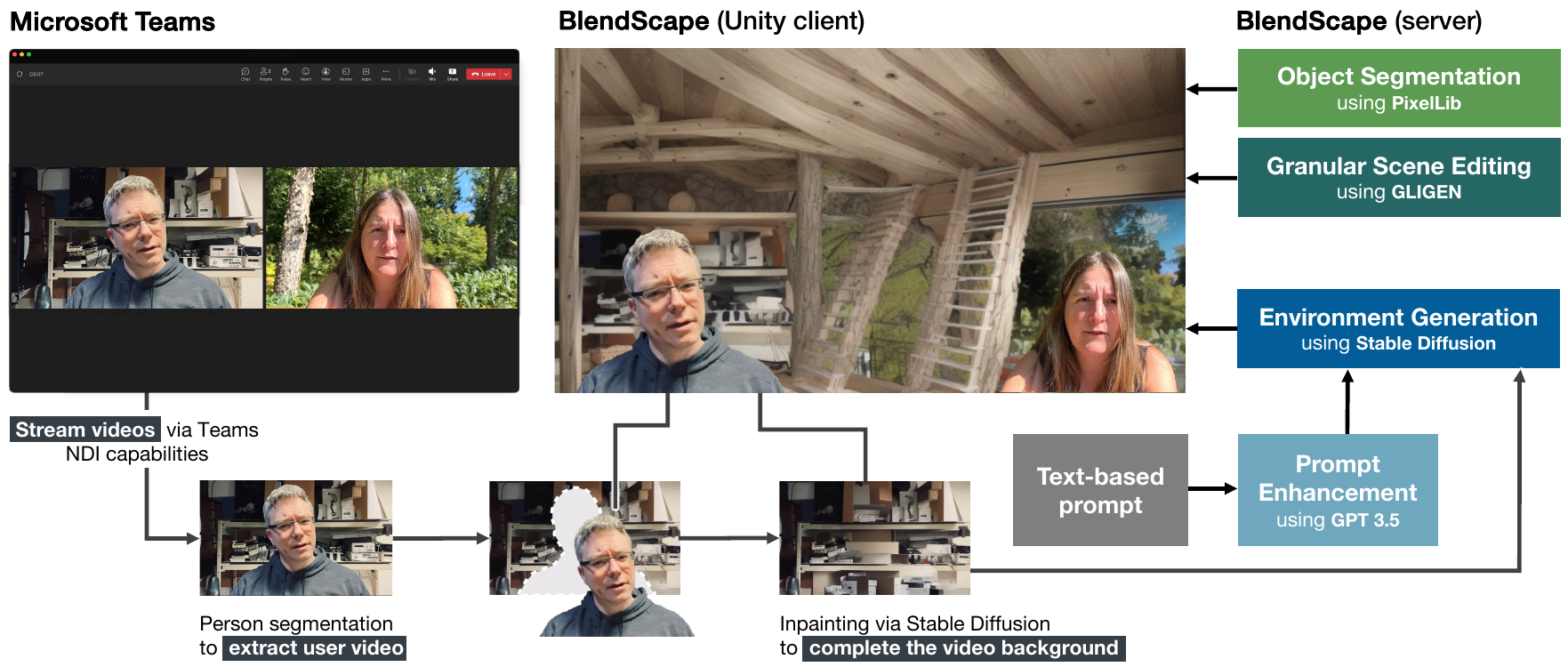}
    \caption{\SystemName Components and Architecture. The \SystemName system consists of a Unity client that serves as the main user interface. Via Microsoft Teams NDI capabilities, the Unity client receives and processes users' videos by performing person segmentation to separate the users from their video backgrounds and using inpainting techniques to fill the missing areas in the backgrounds. The Unity client connects to two servers that run \textit{(1)} Stable Diffusion image generation processes to enable the environment generation techniques; \textit{(2)} PixelLib for object segmentation to generate foreground objects, GPT-3.5 to enhance users' prompts with contextually-relevant keywords, and GLIGEN (grounded text-to-image generation model) for re-generating small portions of the environment.}
    \Description{BlendScape Components and Architecture. The BlendScape system consists of a Unity client that serves as the main user interface. Via Microsoft Teams NDI capabilities, the Unity client receives and processes users' videos by performing person segmentation to separate the users from their video backgrounds and using inpainting techniques to fill the missing areas in the backgrounds. The Unity client connects to two servers that run (1) Stable Diffusion image generation processes to enable the environment generation techniques; (2) PixelLib for object segmentation to generate foreground objects, GPT-3.5 to enhance users' prompts with contextually relevant keywords, and GLIGEN (grounded text-to-image generation model) for re-generating small portions of the environment.}
    \label{fig:system-components}
\end{figure*}

\subsection{Implementation}

Figure~\ref{fig:system-components} shows \SystemName's key components and system architecture.
We implemented \SystemName's user interface as a Unity application integrated with Microsoft Teams.
\SystemName receives video streams from individual meeting participants via Microsoft Teams' NDI streaming capabilities.

\textbf{Image generation models:} To enable the environment generation techniques, we used open-source Stable Diffusion \cite{rombachHighResolutionImageSynthesis2022} and ControlNet \cite{zhangAddingConditionalControl2023} models through the WebUI API\footnote{\textbf{Stable Diffusion WebUI:} \url{https://github.com/AUTOMATIC1111/stable-diffusion-webui}, \textbf{ControlNet Extension:} \url{https://github.com/Mikubill/sd-webui-controlnet}}, hosted on a PC with an AMD EPYC 7742 64-Core Processor and NVIDIA RTX A6000 GPU. 
For \textit{inpainting}, we used the Realistic Vision 2.0\footnote{\textbf{Realistic Vision 2.0}: \url{https://huggingface.co/SG161222/Realistic_Vision_V2.0}} Stable Diffusion checkpoint which is fine-tuned for inpainting, further guided by a ControlNet inpainting model. 
For \textit{image-to-image} generations, we used Realistic Vision's base checkpoint, guided by ControlNet Depth and Canny models to preserve the spatial layout and salient features of the image priors. 
Furthermore, we incorporated GLIGEN~\cite{liGligenOpensetGrounded2023} to allow users to regenerate specific areas of the scene. 
Similar to ControlNet \cite{zhangAddingConditionalControl2023} in its goal of guiding diffusion models, GLIGEN makes region-specific edits based on textual prompts and bounding box coordinates. 

Using these models, the average generation times were 10s for \textit{inpainted} environments, 25s for \textit{image-to-image} environments, and 20s for GLIGEN-enabled edits. This includes the time to enhance users' prompts with additional keywords (via GPT-3.5).

\begin{figure}[h!]
    \centering
    \includegraphics[width=0.8\linewidth]{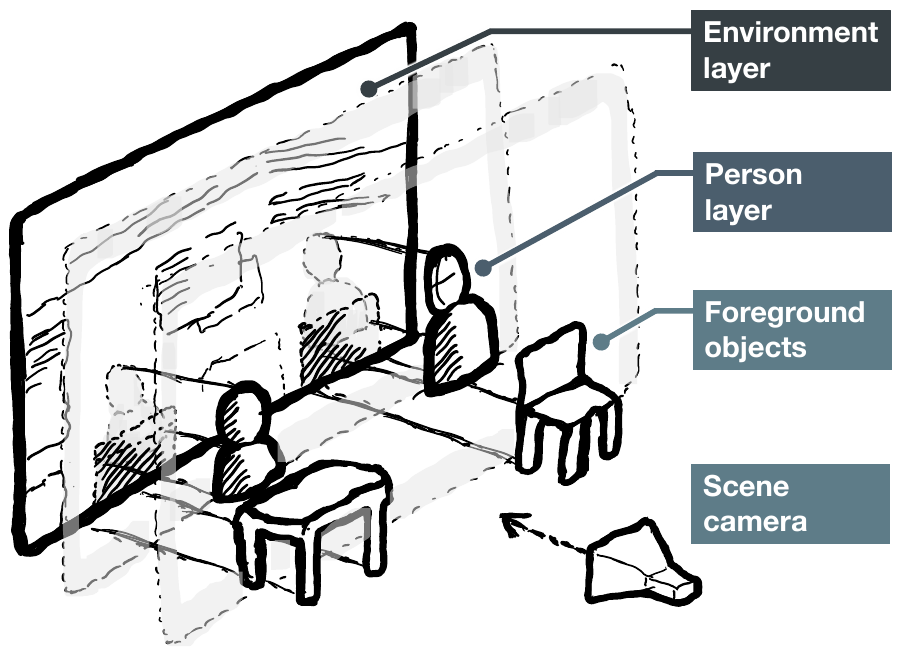}
    \caption{2.5D, Layered Scenes. \SystemName separates the blended environment, users' videos, and foreground objects into layers and renders each layer with an orthographic camera. This enables flexibly combining the layers as input to image generation models to perform \textit{inpainting} and \textit{image-to-image} transformations.}
    \label{fig:system-layers}
    \Description{Decomposition of the BlendScape scene to demonstrate the layered approach. Separate elements of the scene are grouped into layers and subsequently rendered with orthographic cameras, enabling the combination of possibly multiple environment generation passes. Layers include (from back to front): blended background image, person layer, foreground layer.}
\end{figure}

\textbf{2.5D, layered scenes:} 
The \textit{image-to-image} generation mode transforms the entire existing environment, while the \textit{inpainting} mode only takes users' webcam backgrounds as input.
To isolate the correct scene elements for the image generation models, we implemented \SystemName as a 2.5D scene in Unity with five 2D layers staggered at different depths (Fig.~\ref{fig:system-layers}): \textit{(1)}~foreground objects, \textit{(2)}~users' videos (separated from their video backgrounds), \textit{(3)}~users' video backgrounds, \textit{(4)}~background masks (to preserve regions of the video backgrounds), and \textit{(5)}~the generated environment.
We instrumented the Unity scene with multiple cameras that render specific layers, enabling us to capture each layer separately.
We use orthographic cameras (i.e., cameras that do not perform perspective rendering), so scene elements are rendered at a consistent scale even when placed at different depths.

\textbf{Environment segmentation:}
\SystemName uses the PixelLib~\cite{olafenwaSimplifyingObjectSegmentation2021} semantic segmentation model to partition scene elements for hidden surface removal. 
We segment users from their video backgrounds via conventional computer vision techniques.
While \SystemName displays live video feeds of the segmented users, we generate static environments using the first frame of users' videos, due to limitations with running Stable Diffusion at video rates.

Performing person segmentation leaves empty spaces in static video backgrounds, as webcams fail to capture areas of the background that are occluded by the user.
We used Stable Diffusion to \textit{inpaint} this area, simulating a continuous background (Fig.~\ref{fig:system-components}).

\section{Demonstration of Additional Scenarios}
\label{sec:scenario-demonstration}
In this section, we illustrate the expressiveness of \SystemName's generative AI-enabled customization tools through prototyping meeting environments for three distributed collaboration scenarios: \textit{Design Brainstorming}, \textit{Remote Education}, and \textit{Storytime with Family}.

Earlier, we reviewed the range of video-conferencing environments implemented by prior systems to enhance distributed collaborators' communication and sense of co-presence (Sec.~\ref{sec:rel-work-vidconf-envs}).
This review surfaced ten environment design strategies, such as incorporating representations of task spaces, enabling proxemic metaphors, or recording a history of collaborative activities. 
We demonstrate how \SystemName's composition techniques 
can be used to implement eight out of the ten design strategies (shown in bold in Fig.~\ref{fig:rel-work-env-composition}).

Our focus for \SystemName was enabling end-users to compose custom environment visuals; as such, rendering spatial sound was out of scope (Fig.~\ref{fig:rel-work-env-composition}B.iii). 
Currently, \SystemName's 2D image generation techniques do not support rendering egocentric viewing perspectives to simulate face-to-face communication cues (Fig. \ref{fig:rel-work-env-composition}B.v), e.g., turning to face other users seated around a table~\cite{Tang-CSCW23}.
This could be achieved via 360$^{\circ}$ panoramic or 3D environment generation approaches~\cite{Numan-UIST24,  li2023panogen}.

\subsection{Design Brainstorming}
Recall our initial scenario from Sec.~\ref{sec:initial-scenario}, where two interaction designers are remotely collaborating to prototype mixed reality interfaces (Fig.~\ref{fig:scenario1}). 
They use \SystemName to \textbf{render their videos in a unified workspace} that \textbf{blends their physical and digital task spaces} into a single design studio.
This allows them preview their mixed reality designs and simulate face-to-face design critiques, by verbally referring to and gesturing around hand-drawn sketches.

\textbf{Implementation with \SystemName:} We prototyped this scenario using four live video feeds: two webcam feeds of the designers, an external camera feed capturing a physical desk with notebooks, and a screen-capture of a tablet-based digital sketching application (Fig.~\ref{fig:scenario1}a).
We used \SystemName's \textit{inpainting} mode to blend the users' videos and the task space feeds into seamless meeting environments (Fig.~\ref{fig:scenario1}b, c).
Finding the ideal placement of the task space feeds required some trial-and-error; for example, positioning the physical sketches at the bottom of the screen occasionally rendered them on the floor of the environment rather than on a desk.

\begin{figure}[htb!]
    \centering
    \includegraphics[width=\linewidth]{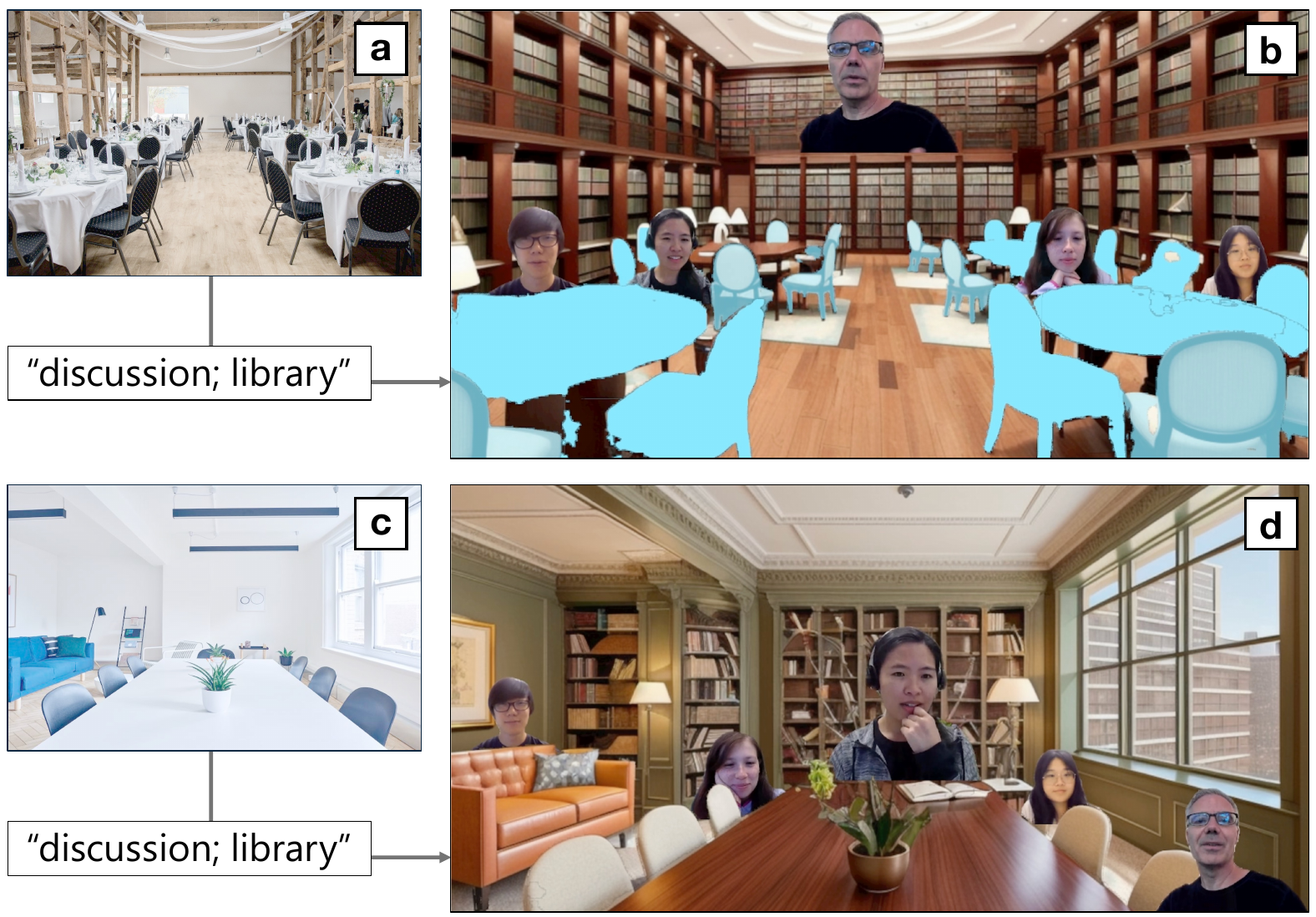}
    \caption{Scenario 2: \textit{Remote Education}. To establish room layouts for seminar discussions, a professor restyles images with small and large tables \textit{(a, b)} to resemble a library setting \textit{(c, d)}.
    This enables students to organize and place themselves behind spatial landmarks in the scene (shown in blue).}
    \Description{Overview of "Remote Education" scenario with the following components and flow. (a, b) Image priors of a wedding venue and conference room. (b) Restyled environment with the wedding venue prior, using a prompt of "library." Chairs and tables are highlighted in blue to indicate objects that users can place themselves behind. (c) Restyled environment with the conference room prior, using a prompt of "library".}
    \label{fig:scenario3}
    \vspace{-1em}
\end{figure}

\subsection{Remote Education}
Next, to demonstrate how \SystemName environments could support spatial metaphors to facilitate conversations~\cite{Hu-CHI23-OpenMic}, we prototyped a scenario involving a remote university course.

\textbf{Scenario Description:}
A professor uses \SystemName to facilitate activities in their literature seminar.
Before the seminar starts, they \textbf{establish the room layout} for small group discussions: they upload an image of a room with several tables and restyle it as a ``\texttt{library}'' (Fig.~\ref{fig:scenario3}a, c).
Students assign themselves to groups by organizing around the tables, leveraging \textbf{proxemic metaphors} (Fig.~\ref{fig:scenario3}c).
To structure a large group discussion later in the seminar, the professor restyles an image of a conference room (Fig.~\ref{fig:scenario3}b, d).
To spotlight speakers, they use \textbf{scaling metaphors} by rendering the students' videos larger and at the head of the table.

\textbf{Implementation with \SystemName:}
We selected two images priors of event spaces that contained enough seating for 5 users and were captured from similar forward-facing perspectives as the users' webcam videos, allowing us to realistically place users behind furniture.
We then used \SystemName's \textit{image-to-image} mode with a \textit{Meeting Theme} prompt of ``\texttt{library}'' to restyle the image priors.

\subsection{Storytime with Family}
Our final scenario is inspired by systems like Waazam~\cite{Hunter-CHI14} and VideoPlay~\cite{follmerVideoPlayPlayful2010}, which mediate playful social interactions between children and family members.

\begin{figure}[htb!]
    \centering
    \includegraphics[width=\linewidth]{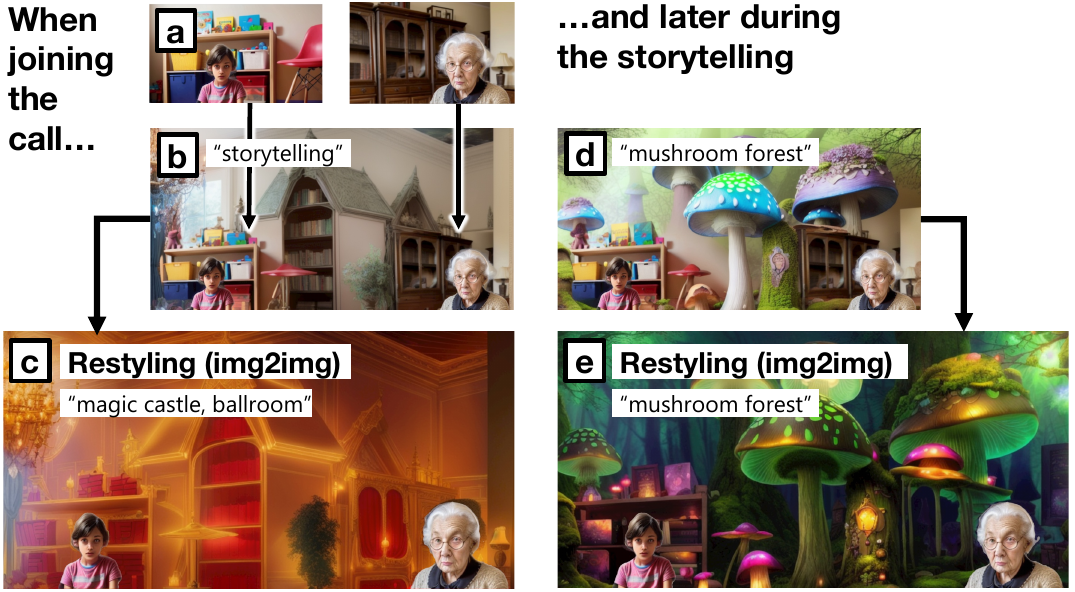}
    \caption{Scenario 3: \textit{Storytime with Family}. A grandmother and her granddaughter use \SystemName to immerse themselves in a fairytale, using \textit{image-to-image} techniques to restyle their blended video backgrounds into a ballroom and mushroom forest.}
    \Description{Overview of "Storytelling" scenario with the following components and flow. (a) Original video feeds. (b) Blended environment with user video feeds composited into the scene based on prompt "storytelling". (c) Restyling of blended "storytelling" environment based on prompt "magic castle, ballroom". (d) New blended environment with user video feeds composited into the scene based on prompt "mushroom forest". (e) Iterative regeneration of blended "mushroom forest" environment.}
    \label{fig:scenario2}
\end{figure}

\textbf{Scenario Description:} A grandmother uses \SystemName to craft a memorable storytelling experience with her granddaughter.
At the start of the call, a castle appears in the space between their video backgrounds with a prompt of ``\texttt{storytelling}'' (Fig.~\ref{fig:scenario2}b).
As the grandmother reads a fairy tale, she restyles the scenes to \textbf{set the theme of the meeting}; the environment transforms into magic ballroom and then a mushroom forest (Fig.~\ref{fig:scenario2}c, e). 
These environments illustrating the story's progression serve as an \textbf{artifact produced through their collaboration}.
A few weeks later, the granddaughter wants to write a sequel to the fairy tale.
They use \SystemName's saved environments and video positions to \textbf{replay a history of their collaboration} and extend their previous scenes.

\textbf{Implementation with \SystemName:} We used the \textit{inpainting} mode to blend the users' physical environments together. We used \textit{image-to-image} techniques with prompts of ``\texttt{magic castle, ballroom}'' and ``\texttt{mushroom forest}'' to restyle the blended environments to reflect the events of the story.

\section{Exploratory Study with End-Users}
\label{sec:study}
To investigate whether and how end-users would find value in using generative AI to personalize video-conferencing environments, we conducted an exploratory study with 15 frequent users of traditional video-conferencing tools (e.g., Microsoft Teams, Zoom).
Our goals were to \textit{(1)}~surface participants' preferences for customizing meeting environments, including visual or layout characteristics they aimed to achieve; \textit{(2)}~investigate to what extent \SystemName allowed participants to express their design intentions.

The study was approved under the Institutional Review Board (Ethics / IRB ID: \#10764; Release and Compliance ID: \#6755).
We conducted 1-hour study sessions remotely via Microsoft Teams.
Participants were compensated with \$50 gift cards. 

\subsection{Participants \& Recruitment}
Advertising through internal mailing lists, we recruited individuals from our organization who use video conferencing tools at least three times per week.
15 individuals participated in our study (7 women, 8 men, majority in an age range of 25-34 years) and
held a variety of job roles: UX or Product Designer (5), UX Researcher (5), Product Manager (3), Software Engineer (1), and Human Factors Engineer (1).
7 out of 15 participants used image generation models once a week or more frequently; the remaining 8 used them once a month or less frequently. 
14 of 15 participants were located in North America and one was located in Asia. 

\subsection{Method}
The study consisted of three environment composition tasks using \SystemName and a post-task discussion.
In the first two tasks, we introduced participants to \SystemName's \textit{inpainting} and \textit{image-to-image} techniques, teaching them to generate environments based on pre-defined prompts and assess the quality in a semi-structured interview portion.
This served as training for Task 3, where participants combined \SystemName's composition tools to create a series of environments for a research meeting scenario.

\textbf{Set-up:} We hosted \SystemName on a local machine and gave participants remote control via Microsoft Teams screen-sharing.
For all tasks, we used pre-recorded videos to represent different users (simulated as NDI streams via the OBS Studio NDI Integration Tools\footnote{\textbf{OBS Studio NDI Integration Tools:} https://obsproject.com/forum/resources/obs-ndi-newtek-ndi\texttrademark-integration-into-obs-studio.528/}), in order to achieve relatively consistent environment generations across participants. 
We used a combination of real physical locations and virtual backgrounds to represent a diverse range of image priors.
For Task 3, we integrated participants' webcam feeds into \SystemName so they could experience being immersed in different environments.

\textbf{Task 1: Walkthrough \& Comparison of Inpainting Techniques} \textit{(10 min)}.
First, we explored blended environment designs for a \textit{Vacation Planning} scenario, where two friends are planning a trip to Paris and generate meeting spaces to reflect landmarks they want to visit.
The facilitator illustrated how even basic input (i.e., providing prompts, changing the position and scale of the image priors) can be used to steer the environment generation via \SystemName's \textit{inpainting} techniques.
Afterwards, we asked participants to experiment with different prompts and layouts.

We then showed participants three environments that preserved varying degrees of users' video backgrounds (Fig.~\ref{fig:task1-envs}).
We asked participants to comment on which examples, if any, they could envision using for the \textit{Vacation Planning} scenario with minimal changes. 
For the environments they could not envision using, we asked them to explain 1-2 key issues.

\textbf{Task 2: Walkthrough \& Comparison of Image-to-Image Techniques} \textit{(10 min)}. 
Next, we introduced \SystemName's \textit{image-to-image} generation techniques using a \textit{Game Stream} scenario, where a Minecraft player uses \SystemName to engage viewers their viewers.
We started from an image of an arcade and used prompts to restyle the image prior in a Minecraft theme. 
Then, we combined \textit{inpainting} and \textit{image-to-image} techniques to blend the users' webcam backgrounds into the arcade image.
Similarly to the first task, we asked participants to compare four example environments (Fig.~\ref{fig:task2-envs}), comment on which options they could envision using for the scenario, and explain key issues they observed. 
We intentionally included examples of roughly blended and cluttered environments to provoke discussions around the limitations of \SystemName's composition techniques.

\textbf{Task 3: Designing Environments for a Progressive Meeting Scenario} \textit{(15 min)}.
To explore how participants could envision adapting environment designs during a live meeting, we instructed them to compose a series of scenes for a scenario involving writing a research paper.
First, we introduced a student character (using a pre-recorded video) and brought the participants' webcam feeds into \SystemName, to play the role of another student. 

We facilitated the task via the following prompts:
\textit{(1)} Design an initial environment for two students to discuss the introduction of the paper; \textit{(2)} The professor joins the call to provide feedback on the students’ ideas. How would you redesign the environment to include them? \textit{(3)} All three users are starting to feel stressed with the paper deadline approaching. How would you redesign the environment to support them?

Throughout the task, we asked the participants to think aloud to describe their design goals and assess the quality of the generated environments. 
We prompted them to use features that we had not yet explored (e.g., adding or removing content).

\textbf{Discussion}  \textit{(15 min)}. We ended with a semi-structured interview around the potential value that generative AI-enabled custom environments could provide to distributed collaboration scenarios. 
We asked participants to comment on specific scenarios where they could or could not envision using a system like \SystemName to personalize meeting environments.
Participants also reflected on unexpected or surprising elements of the environments they generated in the previous tasks and how these aspects might support or detract from collaborative processes.

\begin{figure*}[ht!]
    \centering
    \includegraphics[width=0.9\linewidth]{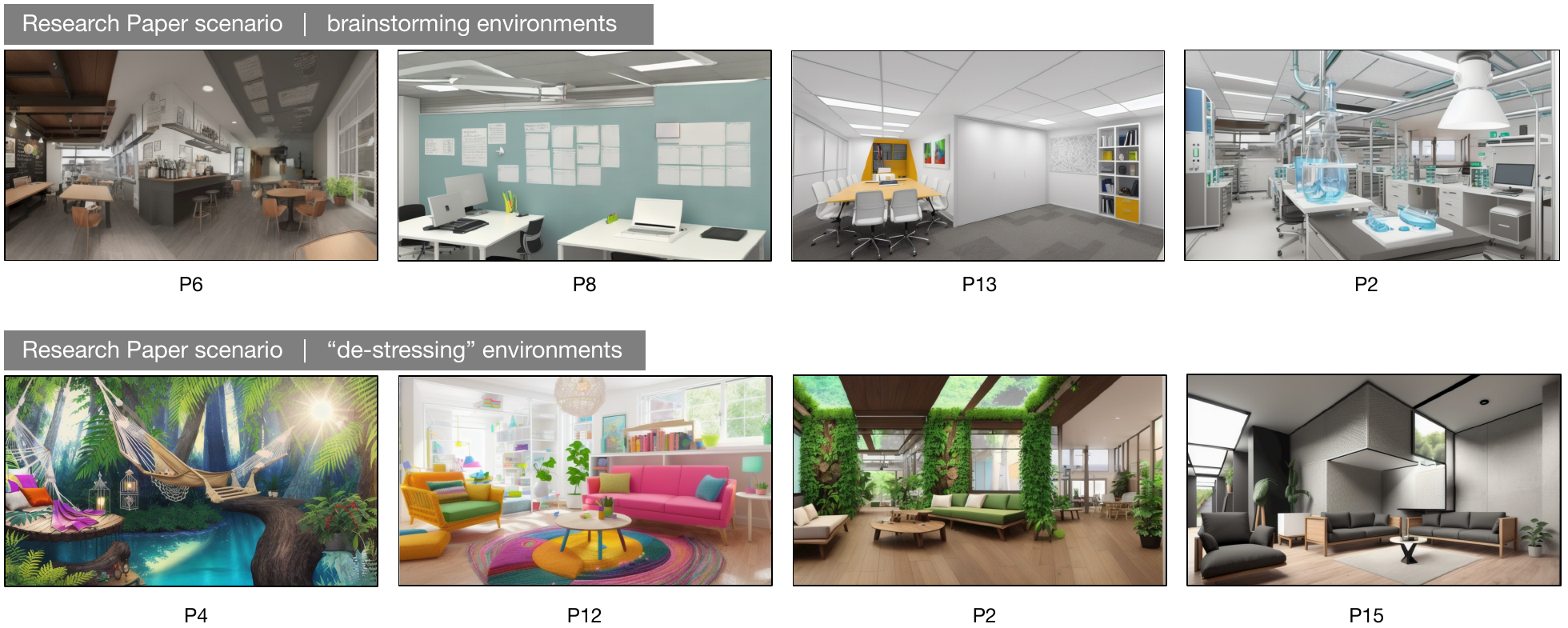}
    \caption{Participants' environment designs for the Research Paper Scenario. To support the students' and professor's writing process in our Task 3 scenario, participants adopted a variety of design strategies: blending their backgrounds into spaces that support creativity (e.g., P6's coffee shop and P8's office space), creating themed environments according to the research topic (e.g., P2's chemistry lab), and removing content to create distraction-free spaces (P13). 
    The participants envisioned both playful and calming environments to ``de-stress'' the users as they approached their paper deadline, e.g., P4's rainforest with hammocks, P12's stuffed animal-themed room, and P2 \& P15's indoor spaces with plants and natural lighting.}
    \Description{Overview of participant environment designs for the Research Paper Scenario. To support the students' and professor's writing process in our Task 3 scenario, participants adopted a variety of design strategies: blending their backgrounds into spaces that support creativity (e.g., P6's coffee shop and P8's office space), creating environments themed according to the research topic (e.g., P2's chemistry lab), and removing content to create empty, distraction-free spaces (P13). The participants envisioned both playful and calming environments to ``de-stress'' the users as they approached their paper deadline, e.g., P4's rainforest with hammocks, P12's stuffed animal-themed room, and P2 \& P15's indoor spaces with plants and natural lighting.}
    \label{fig:results-sample}
\end{figure*}

\subsection{Data Collection \& Analysis}
For all study sessions, we collected audio transcripts \& recordings, screen recordings, and images of the environments generated with \SystemName for later analysis.
To analyze the environment characteristics that participants aimed to achieve, we used a thematic analysis approach~\cite{BraunClarke}. 
One author reviewed all transcripts to create an initial codebook, specifically paying attention to participants' rationale for selecting an environment over another in the Task 1 \& 2 comparison exercises and their revision strategies in Task 3. 
Two authors then reviewed the codes in the context of specific examples provided by the participants (including both quotes and images of the environments) to extract higher-level themes. 
We used an affinity diagramming approach~\cite{Scupin-1997} to analyze and aggregate themes from the post-task discussions around the benefits and limitations of \SystemName and future usage scenarios.

\section{Results}
Figure ~\ref{fig:results-sample} shows a sample of the environments generated by the 15 participants in our study.
In this section, we first present four themes around participants' environment customization preferences (i.e., the visual and layout characteristics they aimed to achieve).
We then discuss the benefits of \SystemName's composition techniques for expressive leverage and limitations with the time and effort required to achieve optimal designs.

\subsection{Environment Customization Preferences}
When assessing the quality of \SystemName's environments, participants expressed preferences for \textit{(1)}~authentic over artificial spaces; \textit{(2)}~both strong and subtle thematic elements, depending on the meeting context; \textit{(3)}~structuring collaboration through spatial layouts of users; \textit{(4)}~balancing the spatial-richness of environments.

\paragraph{\textnormal{\textbf{Authentic environments were preferred over artificial, but came with higher expectations for realism.}}}
A majority of participants expressed that blending users' physical surroundings into a unified meeting space could promote co-presence while maintaining familiar aspects of their individual environments (P2, P5-6, P9, P11-14).
P13 commented that ``taking something personal to [them] and tweaking it'' would help remote collaborators to ``trick [their] minds to believe that [they're] more in the same place.''

\begin{figure}[h!]
    \centering
    \includegraphics[width=\linewidth]{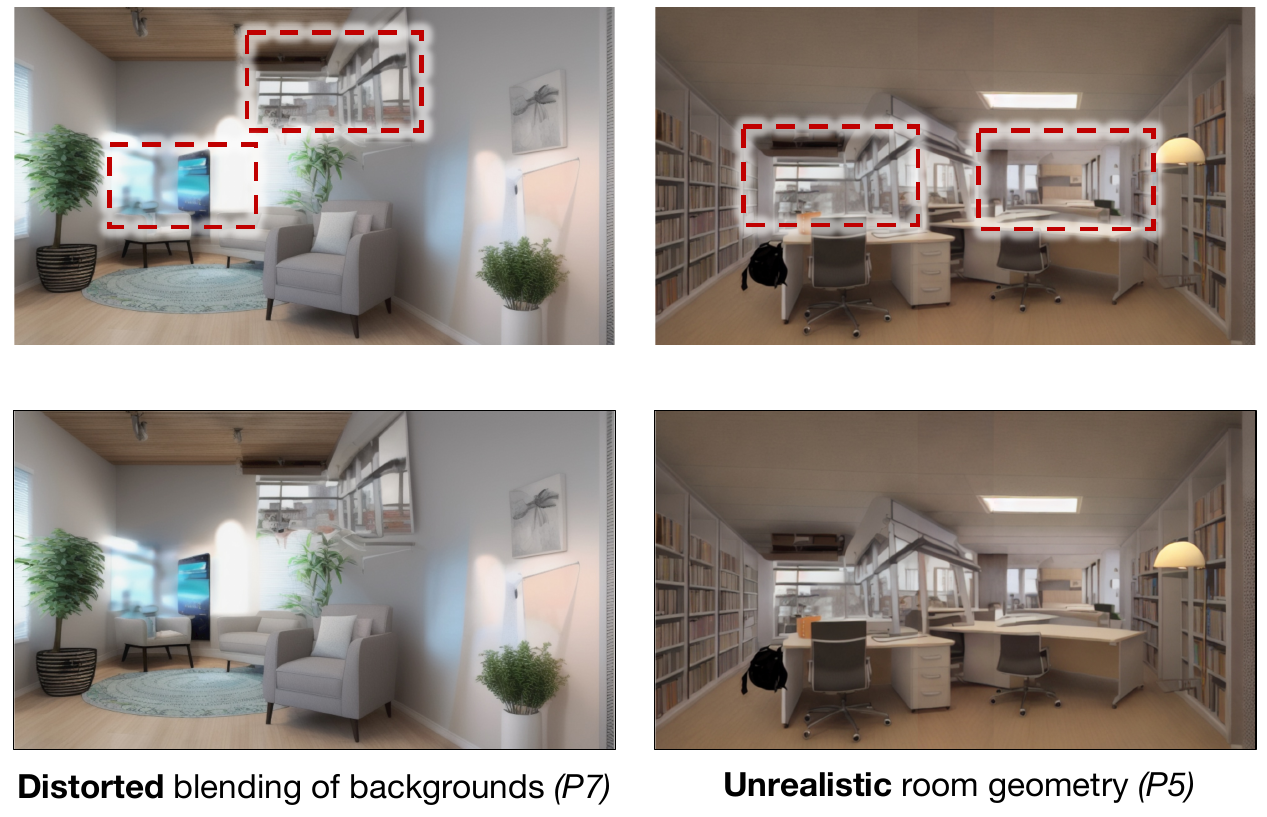}
    \caption{Imperfections in Blended Environments: \SystemName's inpainting techniques can sometimes produce warped or unrealistic room geometries when users' video backgrounds are very disparate (P7) or are captured from different perspectives (P5).}
    \label{fig:results-distorted-backgrounds}
    \Description{Two visual examples of imperfections in blended environments. Firstly, an environment with warped room geometry due to the disparate video backgrounds of the users. Secondly, an environment with warped room geometry due to the different perspectives of the users' video backgrounds.}
\end{figure}

P6 cited another benefit of authentic backgrounds in creating a higher degree of realism, as users' videos appeared to be naturally framed with appropriate furniture and lighting.
In environments constructed from virtual backgrounds, users sometimes appeared to be ``floating in space in front of an image... like a bad Photoshop job'' (P6). 
However, we observed a drawback to this increased realism: participants more easily identified and were more critical of flaws in how their own physical surroundings were blended (P1, P3-5, P7-8, P10), e.g., warped areas or inconsistent room geometry (Fig.~\ref{fig:results-distorted-backgrounds}).
With fully artificial environments, some participants argued that ``there's no pressure to do it well'' (P8); ``even if it was not as polished'' they would have a ``higher tolerance'' for mistakes (P3).

\paragraph{\textnormal{\textbf{Varied preferences for strong vs. subtle visual ties to the meeting context.}}}
A few participants appreciated having stylistic elements that closely reflect the \textit{Meeting Activity} and \textit{Theme} prompts they provided, e.g., a chemistry lab for the \textit{Research Paper} scenario in Task 3.
These participants embraced the at-times unrealistic visuals, arguing ``that we don't need the space to look like traditional meeting spaces'' (P7) which are ``fixed and static... not really imaginative'' (P15).
However, most participants preferred simple and subtle theming to avoid distractions (P1, P7, P9, P13-15).
As P1 stated, environments should ``add texture to the call without pulling away from it.''

\paragraph{\textnormal{\textbf{Establishing user layouts through the environment was a popular design strategy, but required manual staging.}}}
Many participants placed users around spatial landmarks in the scene to reflect specific collaborative activities (P1, P3-7, P9, P13-15).
For example, in the \textit{Research Paper} scenario, P14 positioned the professor in the lower left corner ``to help structure the collaboration,'' similar to a picture-in-picture style for online lectures.
P9 placed users in chairs that were close together to ``give [them] the most privacy'' while having a one-on-one discussion. 

However, some participants observed that \SystemName did not generate spaces to accommodate the existing layout of users in the scene.
P1 and P6 had to manually reposition users to seat them in chairs, but they expected \SystemName to automatically generate chairs and tables to frame the users. 
This points to a limitation of how \SystemName blends environments after segmenting users from their video backgrounds, which we further discuss in Sec.~\ref{sec:disc-composition-techniques}.

\paragraph{\textnormal{\textbf{Need to balance the spatial richness of environments with 2D representations of users.}}}
Participants found that spatial properties of \SystemName's environments (e.g., furniture layouts and lighting aligned with the geometry of the space) increased realism and lent themselves to structuring user layouts for collaboration (as discussed in the previous theme).
However, many participants observed a juxtaposition between these spatially-rich environments and the strictly 2D representation of users (P5-6, P8-10, P11, P13-14).
A few participants tried to tilt the users' videos to ``echo the [3D] perspective in the room'' (P14) and make it look like users were facing each other (P9, P13).
To make our hidden surface removal technique (i.e., hiding floating heads behind foreground objects) more convincing, P14 suggested to ``accept the flatness'' and use environments where all furniture and users' videos are facing forwards, in the style of Teams TogetherMode\footnotemark[1].

\subsection{Benefits and Limitations of Composition Techniques}
We discuss two additional themes around participants' perceptions of the benefits and challenges with using \SystemName to compose meeting environments.

\paragraph{\textnormal{\textbf{\SystemName provides expressive controls for creating environments that reflect the meeting context.}}}
Participants appreciated the ability to rapidly generate and iterate on environment designs with \SystemName, creating over 300 scenes across all three study tasks. 
As P13 expressed, they exerted ``such minimal effort'' to specify their intentions to \SystemName, and the system ``does so much'' to translate their prompts into rich environments. 

Participants' design strategies utilized the full range of our multimodal techniques to steer image generation.
To prototype environments for the \textit{Research Paper} scenario, they added objects that represent the users' research topic (P2-3), removed objects to promote distraction-free collaboration (P10, P13, P14), and prompted for environments they associate with creativity, e.g., parks and coffee shops
(P1, P5-9, P11-12, P15). 
To help ``de-stress'' the students as their deadline approached, a popular strategy was transporting the users to relaxing or playful locations, e.g., a beach, a rainforest, and a stuffed animal-themed space (Fig.~\ref{fig:results-sample}). 
Participants also used the \textit{image-to-image} mode to subtly restyle the existing scene (e.g., using prompts of ``\texttt{warm}'' and ``\texttt{relaxing}'').

\paragraph{\textnormal{\textbf{Challenges with distracting scene elements, time \& effort required to achieve the ideal design.}}}
To enable using a system like \SystemName for professional contexts, participants expressed a need to prevent unexpected elements ``that could prove to be more distracting than helpful'' (P6).
These ranged from minor mistakes that ``look good when you first glance at it'' (P5) but become more apparent upon closer inspection (e.g., two Eiffel Towers in Fig.~\ref{fig:task1-envs}), to more significant cases where the generative AI models misunderstood participants' intent (e.g., GLIGEN inserting a playground slide rather than presentation slides for P9).

While each environment generation takes minimal effort, fixing distracting elements required several iterations (P1, P3-5, P7, P10, P12, P15). 
This impacted when and for which scenarios participants envisioned using generative AI to personalize meeting environments.
For professional scenarios, some participants would prefer to customize meetings spaces before, rather than during, live calls (P1-3, P5), as the ``the uncertainty of the visualizations might detract'' from work discussions (P11).
Scenarios with time pressure may also call for pre-meeting customizations: P12 expressed that their ``therapist charges by the minute, so I don’t want to spend time doing this during a session.''

\section{Discussion}
The scenarios we prototyped with \SystemName (Sec.~\ref{sec:scenario-demonstration}) and our exploratory user study (Sec.~\ref{sec:study}) demonstrated our system's ability to enable a wide range of meeting environment designs.
Overall, all participants expressed they could envision using AI image generation techniques to customize video-conferencing environments in the future, given further controls to prevent unrealistic visuals.

However, further research is required to operationalize similar customization techniques in live video-conferencing systems and understand how to effectively utilize them to achieve different collaboration goals.
We discuss two avenues extending our work with \SystemName: \textit{(1)}~studying the new affordances for distributed collaboration that generative AI-enabled environment composition techniques could enable; \textit{(2)}~technical improvements to address limitations of current-day image generation models.

\subsection{New Opportunities to Facilitate Collaboration through Generative AI-enabled Environment Customization}
While using our classification of prior work's environment design strategies to guide the implementation of \SystemName (Sec.~\ref{sec:rel-work-vidconf-envs}), we observed new ways for these design strategies to manifest when using generative AI.
For example, \SystemName creates literal representations of spatial landmarks (e.g., chairs to establish user layouts, walls to designate boundaries) as opposed to abstract proxemic metaphors in other tools (e.g., the Virtual Floor in OpenMic~\cite{Hu-CHI23-OpenMic}).
Additionally, \SystemName enables ambient representations of task spaces (e.g,. rendering physical sketches in the corner of the screen in our \textit{Design Brainstorming} scenario), as opposed to traditional screen-sharing capabilities which fully direct users' attention towards shared content.

In future work, we would find it interesting to deploy \SystemName in live meeting scenarios to study the impact of these different environment designs on users' collaboration processes, as compared to manually-crafted spaces from prior systems.

We also envision extensions to \SystemName's implementation to enable asymmetric environments and system-driven adaptations.
Per-user, rather than global, customization controls could allow users to tailor meeting spaces to their personal needs (e.g., some users may prefer distraction-free spaces, while others may focus better with visual stimuli in the environment). 
However, these asymmetric environments should be carefully designed to maintain consistency across users when required for the collaboration scenario (e.g., preserving user layouts for turn-taking).
Incorporating real-time summarization of meeting transcripts~\cite{Chen-MeetScript-CSCW23} could enable making the environment a more active partner in collaboration, e.g., automatically adjusting user layouts to transition to new activities or foreshadowing upcoming topics with related visuals.

\subsection{Improving Upon \SystemName's Composition Techniques}
\label{sec:disc-composition-techniques}
We developed \SystemName to achieve a balance between usability requirements for video-conferencing and technical constraints of current image generation models. 
At the time of research, it was infeasible to run image generation models at video rates; as such, we generate environments from a single frame of users' video backgrounds.
However, our approach still enables users to rapidly update environments (via text-based prompts and direct manipulation of video feeds) and render dynamic content by blending in external camera feeds (as demonstrated in our \textit{Design Brainstorming} scenario, which incorporates a live feed of a physical workspace).

While we expect model performance and output quality to improve in the future, we propose extensions to \SystemName to address two key limitations that were surfaced in our study.

\textbf{\textit{(1)} \SystemName can produce environments that ignore the number and existing placement of users in the scene.}
Currently, \SystemName segments and removes users from their video backgrounds before sending the backgrounds as input to Stable Diffusion (Fig.~\ref{fig:system-components}).
This was an intentional choice to avoid generating unrealistic depictions of humans, which is a known issue with image generation models. 
However, this can result in environments that do not reflect the existing spatial layout of users in the scene (e.g., not containing enough furniture to seat all users).

To achieve more ``people-informed'' compositions, 
we envision using OpenPose models\footnote{\textbf{ControlNet OpenPose:} https://huggingface.co/lllyasviel/sd-controlnet-openpose}, which perform human pose estimation, to generate furniture layouts aligned with users in the scene. 
This approach would likely require multiple passes: first, an \textit{inpainting} step to generate the environment, using input images with users present to support the OpenPose model.
A second \textit{inpainting} pass could detect and correct any distorted representations of people.

\textbf{\textit{(2)} The environment generation techniques can be overwhelmed when users' backgrounds and text-based prompts are in contrast.}
Our study participants noticed that when users' ``backgrounds are so disparate, \SystemName really had trouble integrating them into a believable environment'' (P13). 
We observed this in particular with Task 2 (Fig.~\ref{fig:task2-envs}), where the four video streams had semantically different backgrounds (e.g., medieval castle vs. office space), with little relation to the prompt of \textit{Game Streaming}.

To achieve more cohesive blends, one solution is to mask a larger proportion of users' video backgrounds that are relevant to the meeting prompts, thereby prioritizing these backgrounds in the resulting environment. 
This relevance metric could be computed by comparing the similarity between the text-based prompts and semantic descriptions of the video backgrounds.

\section{Limitations}

\textbf{Lack of baseline comparison:}
Given that \SystemName introduces a new way of composing video-conferencing environments through real-time generative AI techniques, we first sought to understand the potential value that the system could provide to users through an exploratory study.
The meeting scenarios we prototyped (Sec.~\ref{sec:scenario-demonstration}) provide an initial comparison of \SystemName to existing customization suites (e.g., Together Mode, Ohyay\footnotemark[1]), as we demonstrated that \SystemName can be used to implement a majority of design strategies offered by these tools.
While conducting a controlled baseline comparison was out of scope for our work, it would be a promising avenue for future research to understand requirements for manual vs. automated authoring workflows.

\textbf{Generalizability of results to live meeting scenarios:} 
Some aspects of our study design intended to ensure consistency across participants and support their agency during the study may limit the generalizability of our results to real meeting scenarios.
In Tasks 1 \& 2, we used pre-recorded videos with static backgrounds to generate similar environments for all participants, enabling us to compare their feedback and extract themes.
However, dynamic video backgrounds could have elicited different design strategies from our participants (e.g., incorporating physical context changes to give collaborators awareness of each others' background activities). 
In Task 3, most participants chose to use virtual rather than real-world backgrounds, due to their physical surroundings being uninteresting or to preserve privacy. 
We find it encouraging that these participants still brainstormed a variety of scenarios where incorporating authentic environments could provide value, e.g., to connect family members or encourage co-workers to ``share their lives in different ways'' (P9).
Finally, we simulated multi-user scenarios via pre-recorded videos and prompted participants to envision how generative AI techniques could be used to support collaboration.
Future studies with multiple participants should be conducted to surface design strategies and challenges specific to collaborative environment composition.

\textbf{Study sample and novelty effects:} 
Our participants' insights may not generalize to all future users of systems like \SystemName, as we primarily studied with individuals from a UX design and product management background.
However, our participants' diverse experience using image generation models suggests that \SystemName's composition techniques are still accessible to novice users.

Considering the novelty of generative AI techniques, our studies around \SystemName are subject to participant response bias~\cite{Dell-CHI12}. We reduced potential bias by discussing both the benefits and limitations of \SystemName with participants and probing them with examples of poorly designed environments (e.g., in Fig.~\ref{fig:task2-envs}) to elicit critiques and improvements to our system.

\section{Conclusion}
We presented \SystemName, a rendering and composition system for video-conferencing participants to tailor meeting environments to their collaboration contexts, by leveraging AI image generation techniques. 
Through implementing scenarios and conducting evaluations with 15 end-users, we demonstrated a rich set of meeting spaces that \SystemName can enable.
Participants' feedback was encouraging in that they could rapidly express their design intentions with \SystemName and envisioned using similar techniques to facilitate meetings in the future.
Based on their feedback, we proposed improvements to \SystemName to mitigate challenges with incohesive or distracting scene elements.
Future studies around \SystemName could explore the impact of its customization techniques on users' collaborative processes and system-driven approaches to adapting environments (e.g., facilitating activity transitions with visuals representing future topics).

\begin{acks}
We thank Amit Gulati \& Henrik Turbell for their valuable feedback, Sasa Junuzovic \& Pat Sweeney for their advice on working with NDI streams, and Michael Nebeling \& Janet Johnson for their support throughout the project. 
We thank our Microsoft Research colleagues and fellow interns who participated in our video and figures.
Finally, we express our gratitude to our study participants for their time.
\end{acks}

\bibliographystyle{ACM-Reference-Format}
\bibliography{References/video-conferencing, References/xr-collaboration, References/other}


\begin{thebibliography}{66}


\ifx \showCODEN    \undefined \def \showCODEN     #1{\unskip}     \fi
\ifx \showDOI      \undefined \def \showDOI       #1{#1}\fi
\ifx \showISBNx    \undefined \def \showISBNx     #1{\unskip}     \fi
\ifx \showISBNxiii \undefined \def \showISBNxiii  #1{\unskip}     \fi
\ifx \showISSN     \undefined \def \showISSN      #1{\unskip}     \fi
\ifx \showLCCN     \undefined \def \showLCCN      #1{\unskip}     \fi
\ifx \shownote     \undefined \def \shownote      #1{#1}          \fi
\ifx \showarticletitle \undefined \def \showarticletitle #1{#1}   \fi
\ifx \showURL      \undefined \def \showURL       {\relax}        \fi
\providecommand\bibfield[2]{#2}
\providecommand\bibinfo[2]{#2}
\providecommand\natexlab[1]{#1}
\providecommand\showeprint[2][]{arXiv:#2}

\bibitem[\protect\citeauthoryear{Agarwala, Dontcheva, Agrawala, Drucker,
  Colburn, Curless, Salesin, and Cohen}{Agarwala et~al\mbox{.}}{2004}]%
        {agarwalaInteractiveDigitalPhotomontage2004}
\bibfield{author}{\bibinfo{person}{Aseem Agarwala}, \bibinfo{person}{Mira
  Dontcheva}, \bibinfo{person}{Maneesh Agrawala}, \bibinfo{person}{Steven
  Drucker}, \bibinfo{person}{Alex Colburn}, \bibinfo{person}{Brian Curless},
  \bibinfo{person}{David Salesin}, {and} \bibinfo{person}{Michael Cohen}.}
  \bibinfo{year}{2004}\natexlab{}.
\newblock \showarticletitle{Interactive digital photomontage}.
\newblock \bibinfo{journal}{\emph{ACM Transactions on Graphics}}
  \bibinfo{volume}{23}, \bibinfo{number}{3} (\bibinfo{date}{Aug.}
  \bibinfo{year}{2004}), \bibinfo{pages}{294--302}.
\newblock
\showISSN{0730-0301, 1557-7368}
\urldef\tempurl%
\url{https://doi.org/10.1145/1015706.1015718}
\showDOI{\tempurl}


\bibitem[\protect\citeauthoryear{Avidan and Shamir}{Avidan and Shamir}{2007}]%
        {avidanSeamCarvingContentaware2007}
\bibfield{author}{\bibinfo{person}{Shai Avidan} {and} \bibinfo{person}{Ariel
  Shamir}.} \bibinfo{year}{2007}\natexlab{}.
\newblock \showarticletitle{Seam Carving for Content-Aware Image Resizing}. In
  \bibinfo{booktitle}{\emph{{{ACM SIGGRAPH}} 2007 Papers}}
  \emph{(\bibinfo{series}{{{SIGGRAPH}} '07})}. \bibinfo{publisher}{{Association
  for Computing Machinery}}, \bibinfo{address}{{New York, NY, USA}},
  \bibinfo{pages}{10--es}.
\newblock
\showISBNx{978-1-4503-7836-9}
\urldef\tempurl%
\url{https://doi.org/10.1145/1275808.1276390}
\showDOI{\tempurl}


\bibitem[\protect\citeauthoryear{Benabdallah, Bourgault, Peek, and
  Jacobs}{Benabdallah et~al\mbox{.}}{2021}]%
        {Benabdallah-CHI21}
\bibfield{author}{\bibinfo{person}{Gabrielle Benabdallah}, \bibinfo{person}{Sam
  Bourgault}, \bibinfo{person}{Nadya Peek}, {and} \bibinfo{person}{Jennifer
  Jacobs}.} \bibinfo{year}{2021}\natexlab{}.
\newblock \showarticletitle{Remote Learners, Home Makers: How Digital
  Fabrication Was Taught Online During a Pandemic}. In
  \bibinfo{booktitle}{\emph{{CHI} '21: {CHI} Conference on Human Factors in
  Computing Systems, Virtual Event / Yokohama, Japan, May 8-13, 2021}},
  \bibfield{editor}{\bibinfo{person}{Yoshifumi Kitamura},
  \bibinfo{person}{Aaron Quigley}, \bibinfo{person}{Katherine Isbister},
  \bibinfo{person}{Takeo Igarashi}, \bibinfo{person}{Pernille Bj{\o}rn}, {and}
  \bibinfo{person}{Steven~Mark Drucker}} (Eds.). \bibinfo{publisher}{{ACM}},
  \bibinfo{pages}{350:1--350:14}.
\newblock
\urldef\tempurl%
\url{https://doi.org/10.1145/3411764.3445450}
\showDOI{\tempurl}


\bibitem[\protect\citeauthoryear{Brade, Wang, Sousa, Oore, and Grossman}{Brade
  et~al\mbox{.}}{2023}]%
        {bradePromptifyTexttoImageGeneration2023}
\bibfield{author}{\bibinfo{person}{Stephen Brade}, \bibinfo{person}{Bryan
  Wang}, \bibinfo{person}{Mauricio Sousa}, \bibinfo{person}{Sageev Oore}, {and}
  \bibinfo{person}{Tovi Grossman}.} \bibinfo{year}{2023}\natexlab{}.
\newblock \showarticletitle{Promptify: {{Text-to-Image Generation}} through
  {{Interactive Prompt Exploration}} with {{Large Language Models}}}. In
  \bibinfo{booktitle}{\emph{Proceedings of the 36th {{Annual ACM Symposium}} on
  {{User Interface Software}} and {{Technology}}}}. \bibinfo{publisher}{{ACM}},
  \bibinfo{address}{{San Francisco , CA , USA}}.
\newblock
\urldef\tempurl%
\url{https://doi.org/10.48550/arXiv.2304.09337}
\showDOI{\tempurl}
\showeprint[arxiv]{2304.09337}~[cs]


\bibitem[\protect\citeauthoryear{Braun and Clarke}{Braun and Clarke}{2006}]%
        {BraunClarke}
\bibfield{author}{\bibinfo{person}{Virginia Braun} {and}
  \bibinfo{person}{Victoria Clarke}.} \bibinfo{year}{2006}\natexlab{}.
\newblock \showarticletitle{Using thematic analysis in psychology}.
\newblock \bibinfo{journal}{\emph{Qualitative Research in Psychology}}
  \bibinfo{volume}{3}, \bibinfo{number}{2} (\bibinfo{year}{2006}),
  \bibinfo{pages}{77--101}.
\newblock


\bibitem[\protect\citeauthoryear{Brown, Mann, Ryder, Subbiah, Kaplan, Dhariwal,
  Neelakantan, Shyam, Sastry, Askell, Agarwal, {Herbert-Voss}, Krueger,
  Henighan, Child, Ramesh, Ziegler, Wu, Winter, Hesse, Chen, Sigler, Litwin,
  Gray, Chess, Clark, Berner, McCandlish, Radford, Sutskever, and Amodei}{Brown
  et~al\mbox{.}}{2020}]%
        {brownLanguageModelsAre2020}
\bibfield{author}{\bibinfo{person}{Tom Brown}, \bibinfo{person}{Benjamin Mann},
  \bibinfo{person}{Nick Ryder}, \bibinfo{person}{Melanie Subbiah},
  \bibinfo{person}{Jared~D Kaplan}, \bibinfo{person}{Prafulla Dhariwal},
  \bibinfo{person}{Arvind Neelakantan}, \bibinfo{person}{Pranav Shyam},
  \bibinfo{person}{Girish Sastry}, \bibinfo{person}{Amanda Askell},
  \bibinfo{person}{Sandhini Agarwal}, \bibinfo{person}{Ariel {Herbert-Voss}},
  \bibinfo{person}{Gretchen Krueger}, \bibinfo{person}{Tom Henighan},
  \bibinfo{person}{Rewon Child}, \bibinfo{person}{Aditya Ramesh},
  \bibinfo{person}{Daniel Ziegler}, \bibinfo{person}{Jeffrey Wu},
  \bibinfo{person}{Clemens Winter}, \bibinfo{person}{Chris Hesse},
  \bibinfo{person}{Mark Chen}, \bibinfo{person}{Eric Sigler},
  \bibinfo{person}{Mateusz Litwin}, \bibinfo{person}{Scott Gray},
  \bibinfo{person}{Benjamin Chess}, \bibinfo{person}{Jack Clark},
  \bibinfo{person}{Christopher Berner}, \bibinfo{person}{Sam McCandlish},
  \bibinfo{person}{Alec Radford}, \bibinfo{person}{Ilya Sutskever}, {and}
  \bibinfo{person}{Dario Amodei}.} \bibinfo{year}{2020}\natexlab{}.
\newblock \showarticletitle{Language {{Models}} Are {{Few-Shot Learners}}}. In
  \bibinfo{booktitle}{\emph{Advances in {{Neural Information Processing
  Systems}}}}, Vol.~\bibinfo{volume}{33}. \bibinfo{publisher}{{Curran
  Associates, Inc.}}, \bibinfo{pages}{1877--1901}.
\newblock


\bibitem[\protect\citeauthoryear{Buxton}{Buxton}{2009}]%
        {Buxton2009}
\bibfield{author}{\bibinfo{person}{Bill Buxton}.}
  \bibinfo{year}{2009}\natexlab{}.
\newblock \bibinfo{booktitle}{\emph{Mediaspace -- Meaningspace --
  Meetingspace}}.
\newblock \bibinfo{publisher}{Springer London}, \bibinfo{address}{London},
  \bibinfo{pages}{217--231}.
\newblock
\showISBNx{978-1-84882-483-6}
\urldef\tempurl%
\url{https://doi.org/10.1007/978-1-84882-483-6_13}
\showDOI{\tempurl}


\bibitem[\protect\citeauthoryear{Cai, Carney, Zada, and Terry}{Cai
  et~al\mbox{.}}{2021}]%
        {Cai-CHI21}
\bibfield{author}{\bibinfo{person}{Carrie~J. Cai}, \bibinfo{person}{Michelle
  Carney}, \bibinfo{person}{Nida Zada}, {and} \bibinfo{person}{Michael Terry}.}
  \bibinfo{year}{2021}\natexlab{}.
\newblock \showarticletitle{Breakdowns and Breakthroughs: Observing Musicians'
  Responses to the {COVID-19} Pandemic}. In \bibinfo{booktitle}{\emph{{CHI}
  '21: {CHI} Conference on Human Factors in Computing Systems, Virtual Event /
  Yokohama, Japan, May 8-13, 2021}},
  \bibfield{editor}{\bibinfo{person}{Yoshifumi Kitamura},
  \bibinfo{person}{Aaron Quigley}, \bibinfo{person}{Katherine Isbister},
  \bibinfo{person}{Takeo Igarashi}, \bibinfo{person}{Pernille Bj{\o}rn}, {and}
  \bibinfo{person}{Steven~Mark Drucker}} (Eds.). \bibinfo{publisher}{{ACM}},
  \bibinfo{pages}{571:1--571:13}.
\newblock
\urldef\tempurl%
\url{https://doi.org/10.1145/3411764.3445192}
\showDOI{\tempurl}


\bibitem[\protect\citeauthoryear{Chen, Li, Liu, Fowler, and Wang}{Chen
  et~al\mbox{.}}{2023}]%
        {Chen-MeetScript-CSCW23}
\bibfield{author}{\bibinfo{person}{Xinyue Chen}, \bibinfo{person}{Shuo Li},
  \bibinfo{person}{Shipeng Liu}, \bibinfo{person}{Robin Fowler}, {and}
  \bibinfo{person}{Xu Wang}.} \bibinfo{year}{2023}\natexlab{}.
\newblock \showarticletitle{MeetScript: Designing Transcript-based Interactions
  to Support Active Participation in Group Video Meetings}.
\newblock \bibinfo{journal}{\emph{Proceedings of the ACM on Human-Computer
  Interaction}}  \bibinfo{volume}{abs/2309.12115} (\bibinfo{year}{2023}).
\newblock
\urldef\tempurl%
\url{https://doi.org/10.48550/ARXIV.2309.12115}
\showDOI{\tempurl}
\showeprint[arXiv]{2309.12115}


\bibitem[\protect\citeauthoryear{Choi and Diehm}{Choi and Diehm}{2021}]%
        {Choi-21-AestheticFlattening}
\bibfield{author}{\bibinfo{person}{Jaz~Hee{-}jeong Choi} {and}
  \bibinfo{person}{Cade Diehm}.} \bibinfo{year}{2021}\natexlab{}.
\newblock \showarticletitle{Aesthetic flattening}.
\newblock \bibinfo{journal}{\emph{Interactions}} \bibinfo{volume}{28},
  \bibinfo{number}{4} (\bibinfo{year}{2021}), \bibinfo{pages}{21--23}.
\newblock
\urldef\tempurl%
\url{https://doi.org/10.1145/3468080}
\showDOI{\tempurl}


\bibitem[\protect\citeauthoryear{Chung and Adar}{Chung and Adar}{2023a}]%
        {chungArtinterAIpoweredBoundary2023}
\bibfield{author}{\bibinfo{person}{John Joon~Young Chung} {and}
  \bibinfo{person}{Eytan Adar}.} \bibinfo{year}{2023}\natexlab{a}.
\newblock \showarticletitle{Artinter: {{AI-powered Boundary Objects}} for
  {{Commissioning Visual Arts}}}. In \bibinfo{booktitle}{\emph{Proceedings of
  the 2023 {{ACM Designing Interactive Systems Conference}}}}
  \emph{(\bibinfo{series}{{{DIS}} '23})}. \bibinfo{publisher}{{Association for
  Computing Machinery}}, \bibinfo{address}{{New York, NY, USA}},
  \bibinfo{pages}{1997--2018}.
\newblock
\showISBNx{978-1-4503-9893-0}
\urldef\tempurl%
\url{https://doi.org/10.1145/3563657.3595961}
\showDOI{\tempurl}


\bibitem[\protect\citeauthoryear{Chung and Adar}{Chung and Adar}{2023b}]%
        {chungPromptPaintSteeringTexttoImage2023}
\bibfield{author}{\bibinfo{person}{John Joon~Young Chung} {and}
  \bibinfo{person}{Eytan Adar}.} \bibinfo{year}{2023}\natexlab{b}.
\newblock \bibinfo{title}{{{PromptPaint}}: {{Steering Text-to-Image Generation
  Through Paint Medium-like Interactions}}}.
\newblock
\newblock
\urldef\tempurl%
\url{https://doi.org/10.1145/3586183.3606777}
\showDOI{\tempurl}
\showeprint[arxiv]{2308.05184}~[cs]


\bibitem[\protect\citeauthoryear{Chung, Kim, Yoo, Lee, Adar, and Chang}{Chung
  et~al\mbox{.}}{2022}]%
        {chungTaleBrushSketchingStories2022}
\bibfield{author}{\bibinfo{person}{John Joon~Young Chung},
  \bibinfo{person}{Wooseok Kim}, \bibinfo{person}{Kang~Min Yoo},
  \bibinfo{person}{Hwaran Lee}, \bibinfo{person}{Eytan Adar}, {and}
  \bibinfo{person}{Minsuk Chang}.} \bibinfo{year}{2022}\natexlab{}.
\newblock \showarticletitle{{{TaleBrush}}: {{Sketching Stories}} with
  {{Generative Pretrained Language Models}}}. In
  \bibinfo{booktitle}{\emph{{{CHI Conference}} on {{Human Factors}} in
  {{Computing Systems}}}}. \bibinfo{publisher}{{ACM}}, \bibinfo{address}{{New
  Orleans LA USA}}, \bibinfo{pages}{1--19}.
\newblock
\showISBNx{978-1-4503-9157-3}
\urldef\tempurl%
\url{https://doi.org/10.1145/3491102.3501819}
\showDOI{\tempurl}


\bibitem[\protect\citeauthoryear{Coyne and Sproat}{Coyne and Sproat}{2001}]%
        {coyneWordsEyeAutomaticTexttoscene2001}
\bibfield{author}{\bibinfo{person}{Bob Coyne} {and} \bibinfo{person}{Richard
  Sproat}.} \bibinfo{year}{2001}\natexlab{}.
\newblock \showarticletitle{{{WordsEye}}: An Automatic Text-to-Scene Conversion
  System}. In \bibinfo{booktitle}{\emph{Proceedings of the 28th Annual
  Conference on {{Computer}} Graphics and Interactive Techniques}}
  \emph{(\bibinfo{series}{{{SIGGRAPH}} '01})}. \bibinfo{publisher}{{Association
  for Computing Machinery}}, \bibinfo{address}{{New York, NY, USA}},
  \bibinfo{pages}{487--496}.
\newblock
\showISBNx{978-1-58113-374-5}
\urldef\tempurl%
\url{https://doi.org/10.1145/383259.383316}
\showDOI{\tempurl}


\bibitem[\protect\citeauthoryear{Dang, Brudy, Fitzmaurice, and Anderson}{Dang
  et~al\mbox{.}}{2023}]%
        {dangWorldSmithIterativeExpressive2023}
\bibfield{author}{\bibinfo{person}{Hai Dang}, \bibinfo{person}{Frederik Brudy},
  \bibinfo{person}{George Fitzmaurice}, {and} \bibinfo{person}{Fraser
  Anderson}.} \bibinfo{year}{2023}\natexlab{}.
\newblock \bibinfo{title}{{{WorldSmith}}: {{Iterative}} and {{Expressive
  Prompting}} for {{World Building}} with a {{Generative AI}}}.
\newblock
\newblock
\showeprint[arxiv]{2308.13355}~[cs]


\bibitem[\protect\citeauthoryear{Dang, Mecke, and Buschek}{Dang
  et~al\mbox{.}}{2022}]%
        {dangGANSliderHowUsers2022}
\bibfield{author}{\bibinfo{person}{Hai Dang}, \bibinfo{person}{Lukas Mecke},
  {and} \bibinfo{person}{Daniel Buschek}.} \bibinfo{year}{2022}\natexlab{}.
\newblock \showarticletitle{{{GANSlider}}: {{How Users Control Generative
  Models}} for {{Images}} Using {{Multiple Sliders}} with and without
  {{Feedforward Information}}}. In \bibinfo{booktitle}{\emph{Proceedings of the
  2022 {{CHI Conference}} on {{Human Factors}} in {{Computing Systems}}}}
  \emph{(\bibinfo{series}{{{CHI}} '22})}. \bibinfo{publisher}{{Association for
  Computing Machinery}}, \bibinfo{address}{{New York, NY, USA}},
  \bibinfo{pages}{1--15}.
\newblock
\showISBNx{978-1-4503-9157-3}
\urldef\tempurl%
\url{https://doi.org/10.1145/3491102.3502141}
\showDOI{\tempurl}


\bibitem[\protect\citeauthoryear{Dell, Vaidyanathan, Medhi, Cutrell, and
  Thies}{Dell et~al\mbox{.}}{2012}]%
        {Dell-CHI12}
\bibfield{author}{\bibinfo{person}{Nicola Dell}, \bibinfo{person}{Vidya
  Vaidyanathan}, \bibinfo{person}{Indrani Medhi}, \bibinfo{person}{Edward
  Cutrell}, {and} \bibinfo{person}{William Thies}.}
  \bibinfo{year}{2012}\natexlab{}.
\newblock \showarticletitle{"Yours is better!": participant response bias in
  {HCI}}. In \bibinfo{booktitle}{\emph{{CHI} Conference on Human Factors in
  Computing Systems, {CHI} '12, Austin, TX, {USA} - May 05 - 10, 2012}},
  \bibfield{editor}{\bibinfo{person}{Joseph~A. Konstan}, \bibinfo{person}{Ed~H.
  Chi}, {and} \bibinfo{person}{Kristina H{\"{o}}{\"{o}}k}} (Eds.).
  \bibinfo{publisher}{{ACM}}, \bibinfo{pages}{1321--1330}.
\newblock
\urldef\tempurl%
\url{https://doi.org/10.1145/2207676.2208589}
\showDOI{\tempurl}


\bibitem[\protect\citeauthoryear{{El-Nouby}, Sharma, Schulz, Hjelm, Asri,
  Kahou, Bengio, and Taylor}{{El-Nouby} et~al\mbox{.}}{2019}]%
        {el-noubyTellDrawRepeat2019}
\bibfield{author}{\bibinfo{person}{Alaaeldin {El-Nouby}},
  \bibinfo{person}{Shikhar Sharma}, \bibinfo{person}{Hannes Schulz},
  \bibinfo{person}{R~Devon Hjelm}, \bibinfo{person}{Layla~El Asri},
  \bibinfo{person}{Samira~Ebrahimi Kahou}, \bibinfo{person}{Yoshua Bengio},
  {and} \bibinfo{person}{Graham Taylor}.} \bibinfo{year}{2019}\natexlab{}.
\newblock \showarticletitle{Tell, {{Draw}}, and {{Repeat}}: {{Generating}} and
  {{Modifying Images Based}} on {{Continual Linguistic Instruction}}}. In
  \bibinfo{booktitle}{\emph{2019 {{IEEE}}/{{CVF International Conference}} on
  {{Computer Vision}} ({{ICCV}})}}. \bibinfo{publisher}{{IEEE}},
  \bibinfo{address}{{Seoul, Korea (South)}}, \bibinfo{pages}{10303--10311}.
\newblock
\showISBNx{978-1-72814-803-8}
\urldef\tempurl%
\url{https://doi.org/10.1109/ICCV.2019.01040}
\showDOI{\tempurl}


\bibitem[\protect\citeauthoryear{Fauville, Luo, Queiroz, Bailenson, and
  Hancock}{Fauville et~al\mbox{.}}{2021}]%
        {Fauville-2021-ZoomExhaustion}
\bibfield{author}{\bibinfo{person}{G. Fauville}, \bibinfo{person}{M. Luo},
  \bibinfo{person}{A.C.M. Queiroz}, \bibinfo{person}{J.N. Bailenson}, {and}
  \bibinfo{person}{J. Hancock}.} \bibinfo{year}{2021}\natexlab{}.
\newblock \showarticletitle{Zoom Exhaustion \& Fatigue Scale}.
\newblock \bibinfo{journal}{\emph{Computers in Human Behavior Reports}}
  \bibinfo{volume}{4} (\bibinfo{year}{2021}), \bibinfo{pages}{100119}.
\newblock
\showISSN{2451-9588}
\urldef\tempurl%
\url{https://doi.org/10.1016/j.chbr.2021.100119}
\showDOI{\tempurl}


\bibitem[\protect\citeauthoryear{Follmer, Raffle, Go, Ballagas, and
  Ishii}{Follmer et~al\mbox{.}}{2010}]%
        {follmerVideoPlayPlayful2010}
\bibfield{author}{\bibinfo{person}{Sean Follmer}, \bibinfo{person}{Hayes
  Raffle}, \bibinfo{person}{Janet Go}, \bibinfo{person}{Rafael Ballagas}, {and}
  \bibinfo{person}{Hiroshi Ishii}.} \bibinfo{year}{2010}\natexlab{}.
\newblock \showarticletitle{Video Play: Playful Interactions in Video
  Conferencing for Long-Distance Families with Young Children}. In
  \bibinfo{booktitle}{\emph{Proceedings of the 9th {{International Conference}}
  on {{Interaction Design}} and {{Children}}}}. \bibinfo{publisher}{{ACM}},
  \bibinfo{address}{{Barcelona Spain}}, \bibinfo{pages}{49--58}.
\newblock
\showISBNx{978-1-60558-951-0}
\urldef\tempurl%
\url{https://doi.org/10.1145/1810543.1810550}
\showDOI{\tempurl}


\bibitem[\protect\citeauthoryear{Fuchsberger, Beuthel, Bentegeac, and
  Tscheligi}{Fuchsberger et~al\mbox{.}}{2021}]%
        {Fuchsberger-CHI21}
\bibfield{author}{\bibinfo{person}{Verena Fuchsberger},
  \bibinfo{person}{Janne~Mascha Beuthel}, \bibinfo{person}{Philippe Bentegeac},
  {and} \bibinfo{person}{Manfred Tscheligi}.} \bibinfo{year}{2021}\natexlab{}.
\newblock \showarticletitle{Grandparents and Grandchildren Meeting Online: The
  Role of Material Things in Remote Settings}. In
  \bibinfo{booktitle}{\emph{{CHI} '21: {CHI} Conference on Human Factors in
  Computing Systems, Virtual Event / Yokohama, Japan, May 8-13, 2021}},
  \bibfield{editor}{\bibinfo{person}{Yoshifumi Kitamura},
  \bibinfo{person}{Aaron Quigley}, \bibinfo{person}{Katherine Isbister},
  \bibinfo{person}{Takeo Igarashi}, \bibinfo{person}{Pernille Bj{\o}rn}, {and}
  \bibinfo{person}{Steven~Mark Drucker}} (Eds.). \bibinfo{publisher}{{ACM}},
  \bibinfo{pages}{478:1--478:14}.
\newblock
\urldef\tempurl%
\url{https://doi.org/10.1145/3411764.3445191}
\showDOI{\tempurl}


\bibitem[\protect\citeauthoryear{Gkioxari, Malik, and Johnson}{Gkioxari
  et~al\mbox{.}}{2020}]%
        {gkioxariMeshRCNN2020}
\bibfield{author}{\bibinfo{person}{Georgia Gkioxari}, \bibinfo{person}{Jitendra
  Malik}, {and} \bibinfo{person}{Justin Johnson}.}
  \bibinfo{year}{2020}\natexlab{}.
\newblock \bibinfo{title}{Mesh {{R-CNN}}}.
\newblock
\newblock
\showeprint[arxiv]{1906.02739}~[cs]


\bibitem[\protect\citeauthoryear{Gr{\o}nb{\ae}k, Mackay, Borowski,
  {Beaudouin-Lafon}, Hoggan, and Klokmose}{Gr{\o}nb{\ae}k
  et~al\mbox{.}}{2023a}]%
        {Gronbaek-UIST23}
\bibfield{author}{\bibinfo{person}{Jens~Emil Gr{\o}nb{\ae}k},
  \bibinfo{person}{Wendy~E Mackay}, \bibinfo{person}{Marcel Borowski},
  \bibinfo{person}{Michel {Beaudouin-Lafon}}, \bibinfo{person}{Eve Hoggan},
  {and} \bibinfo{person}{Clemens~N Klokmose}.}
  \bibinfo{year}{2023}\natexlab{a}.
\newblock \showarticletitle{Mirrorverse: {{Live Tailoring}} of {{Video
  Conferencing Interfaces}}}. In \bibinfo{booktitle}{\emph{Proceedings of the
  36th {{Annual ACM Symposium}} on {{User Interface Software}} and
  {{Technology}}}}. \bibinfo{publisher}{{ACM}}, \bibinfo{address}{{San
  Francisco , CA , USA}}.
\newblock
\urldef\tempurl%
\url{https://doi.org/10.1145/3586183.3606767}
\showDOI{\tempurl}


\bibitem[\protect\citeauthoryear{Gr{\o}nb{\ae}k, Saat{\c{c}}i, Griggio, and
  Klokmose}{Gr{\o}nb{\ae}k et~al\mbox{.}}{2021}]%
        {Gronbaek-CHI21}
\bibfield{author}{\bibinfo{person}{Jens~Emil Gr{\o}nb{\ae}k},
  \bibinfo{person}{Banu Saat{\c{c}}i}, \bibinfo{person}{Carla~F. Griggio},
  {and} \bibinfo{person}{Clemens~Nylandsted Klokmose}.}
  \bibinfo{year}{2021}\natexlab{}.
\newblock \showarticletitle{MirrorBlender: Supporting Hybrid Meetings with a
  Malleable Video-Conferencing System}. In \bibinfo{booktitle}{\emph{{CHI} '21:
  {CHI} Conference on Human Factors in Computing Systems, Virtual Event /
  Yokohama, Japan, May 8-13, 2021}},
  \bibfield{editor}{\bibinfo{person}{Yoshifumi Kitamura},
  \bibinfo{person}{Aaron Quigley}, \bibinfo{person}{Katherine Isbister},
  \bibinfo{person}{Takeo Igarashi}, \bibinfo{person}{Pernille Bj{\o}rn}, {and}
  \bibinfo{person}{Steven~Mark Drucker}} (Eds.). \bibinfo{publisher}{{ACM}},
  \bibinfo{pages}{451:1--451:13}.
\newblock
\urldef\tempurl%
\url{https://doi.org/10.1145/3411764.3445698}
\showDOI{\tempurl}


\bibitem[\protect\citeauthoryear{Gr{\o}nb{\ae}k, Pfeuffer, Velloso, Astrup,
  Pedersen, Kj{\ae}r, Leiva, and Gellersen}{Gr{\o}nb{\ae}k
  et~al\mbox{.}}{2023b}]%
        {Gronbaek-CHI23}
\bibfield{author}{\bibinfo{person}{Jens Emil~Sloth Gr{\o}nb{\ae}k},
  \bibinfo{person}{Ken Pfeuffer}, \bibinfo{person}{Eduardo Velloso},
  \bibinfo{person}{Morten Astrup}, \bibinfo{person}{Melanie
  Isabel~S{\o}nderk{\ae}r Pedersen}, \bibinfo{person}{Martin Kj{\ae}r},
  \bibinfo{person}{Germ{\'{a}}n Leiva}, {and} \bibinfo{person}{Hans
  Gellersen}.} \bibinfo{year}{2023}\natexlab{b}.
\newblock \showarticletitle{Partially Blended Realities: Aligning Dissimilar
  Spaces for Distributed Mixed Reality Meetings}. In
  \bibinfo{booktitle}{\emph{Proceedings of the 2023 {CHI} Conference on Human
  Factors in Computing Systems, {CHI} 2023, Hamburg, Germany, April 23-28,
  2023}}, \bibfield{editor}{\bibinfo{person}{Albrecht Schmidt},
  \bibinfo{person}{Kaisa V{\"{a}}{\"{a}}n{\"{a}}nen}, \bibinfo{person}{Tesh
  Goyal}, \bibinfo{person}{Per~Ola Kristensson}, \bibinfo{person}{Anicia
  Peters}, \bibinfo{person}{Stefanie Mueller}, \bibinfo{person}{Julie~R.
  Williamson}, {and} \bibinfo{person}{Max~L. Wilson}} (Eds.).
  \bibinfo{publisher}{{ACM}}, \bibinfo{pages}{456:1--456:16}.
\newblock
\urldef\tempurl%
\url{https://doi.org/10.1145/3544548.3581515}
\showDOI{\tempurl}


\bibitem[\protect\citeauthoryear{Han, Geiskkovitch, Yuan, Mills, Zhong, Chen,
  Stuerzlinger, and Neustaedter}{Han et~al\mbox{.}}{2023}]%
        {Han-CHI23}
\bibfield{author}{\bibinfo{person}{Dongqi Han}, \bibinfo{person}{Denise~Y.
  Geiskkovitch}, \bibinfo{person}{Ye Yuan}, \bibinfo{person}{Chelsea Mills},
  \bibinfo{person}{Ce Zhong}, \bibinfo{person}{Amy Yo~Sue Chen},
  \bibinfo{person}{Wolfgang Stuerzlinger}, {and} \bibinfo{person}{Carman
  Neustaedter}.} \bibinfo{year}{2023}\natexlab{}.
\newblock \showarticletitle{Dr.'s Eye: The Design and Evaluation of a Video
  Conferencing System to Support Doctor Appointments in Home Settings}. In
  \bibinfo{booktitle}{\emph{Proceedings of the 2023 {CHI} Conference on Human
  Factors in Computing Systems, {CHI} 2023, Hamburg, Germany, April 23-28,
  2023}}, \bibfield{editor}{\bibinfo{person}{Albrecht Schmidt},
  \bibinfo{person}{Kaisa V{\"{a}}{\"{a}}n{\"{a}}nen}, \bibinfo{person}{Tesh
  Goyal}, \bibinfo{person}{Per~Ola Kristensson}, \bibinfo{person}{Anicia
  Peters}, \bibinfo{person}{Stefanie Mueller}, \bibinfo{person}{Julie~R.
  Williamson}, {and} \bibinfo{person}{Max~L. Wilson}} (Eds.).
  \bibinfo{publisher}{{ACM}}, \bibinfo{pages}{343:1--343:18}.
\newblock
\urldef\tempurl%
\url{https://doi.org/10.1145/3544548.3581350}
\showDOI{\tempurl}


\bibitem[\protect\citeauthoryear{Herskovitz, Cheng, Guo, Sample, and
  Nebeling}{Herskovitz et~al\mbox{.}}{2022}]%
        {Herskovitz-ISS22}
\bibfield{author}{\bibinfo{person}{Jaylin Herskovitz}, \bibinfo{person}{Yifei
  Cheng}, \bibinfo{person}{Anhong Guo}, \bibinfo{person}{Alanson~P. Sample},
  {and} \bibinfo{person}{Michael Nebeling}.} \bibinfo{year}{2022}\natexlab{}.
\newblock \showarticletitle{XSpace: An Augmented Reality Toolkit for Enabling
  Spatially-Aware Distributed Collaboration}.
\newblock \bibinfo{journal}{\emph{Proc. {ACM} Hum. Comput. Interact.}}
  \bibinfo{volume}{6}, \bibinfo{number}{{ISS}} (\bibinfo{year}{2022}),
  \bibinfo{pages}{277--302}.
\newblock
\urldef\tempurl%
\url{https://doi.org/10.1145/3567721}
\showDOI{\tempurl}


\bibitem[\protect\citeauthoryear{Hu, Gr{\o}nb{\ae}k, Houck, and Heo}{Hu
  et~al\mbox{.}}{2023a}]%
        {Hu-CHI23-OpenMic}
\bibfield{author}{\bibinfo{person}{Erzhen Hu}, \bibinfo{person}{Jens Emil~Sloth
  Gr{\o}nb{\ae}k}, \bibinfo{person}{Austin Houck}, {and}
  \bibinfo{person}{Seongkook Heo}.} \bibinfo{year}{2023}\natexlab{a}.
\newblock \showarticletitle{OpenMic: Utilizing Proxemic Metaphors for
  Conversational Floor Transitions in Multiparty Video Meetings}. In
  \bibinfo{booktitle}{\emph{Proceedings of the 2023 {CHI} Conference on Human
  Factors in Computing Systems, {CHI} 2023, Hamburg, Germany, April 23-28,
  2023}}, \bibfield{editor}{\bibinfo{person}{Albrecht Schmidt},
  \bibinfo{person}{Kaisa V{\"{a}}{\"{a}}n{\"{a}}nen}, \bibinfo{person}{Tesh
  Goyal}, \bibinfo{person}{Per~Ola Kristensson}, \bibinfo{person}{Anicia
  Peters}, \bibinfo{person}{Stefanie Mueller}, \bibinfo{person}{Julie~R.
  Williamson}, {and} \bibinfo{person}{Max~L. Wilson}} (Eds.).
  \bibinfo{publisher}{{ACM}}, \bibinfo{pages}{793:1--793:17}.
\newblock
\urldef\tempurl%
\url{https://doi.org/10.1145/3544548.3581013}
\showDOI{\tempurl}


\bibitem[\protect\citeauthoryear{Hu, Gr{\o}nb{\ae}k, Ying, Du, and Heo}{Hu
  et~al\mbox{.}}{2023b}]%
        {Hu-CHI23-ThingShare}
\bibfield{author}{\bibinfo{person}{Erzhen Hu}, \bibinfo{person}{Jens Emil~Sloth
  Gr{\o}nb{\ae}k}, \bibinfo{person}{Wen Ying}, \bibinfo{person}{Ruofei Du},
  {and} \bibinfo{person}{Seongkook Heo}.} \bibinfo{year}{2023}\natexlab{b}.
\newblock \showarticletitle{ThingShare: Ad-Hoc Digital Copies of Physical
  Objects for Sharing Things in Video Meetings}. In
  \bibinfo{booktitle}{\emph{Proceedings of the 2023 {CHI} Conference on Human
  Factors in Computing Systems, {CHI} 2023, Hamburg, Germany, April 23-28,
  2023}}, \bibfield{editor}{\bibinfo{person}{Albrecht Schmidt},
  \bibinfo{person}{Kaisa V{\"{a}}{\"{a}}n{\"{a}}nen}, \bibinfo{person}{Tesh
  Goyal}, \bibinfo{person}{Per~Ola Kristensson}, \bibinfo{person}{Anicia
  Peters}, \bibinfo{person}{Stefanie Mueller}, \bibinfo{person}{Julie~R.
  Williamson}, {and} \bibinfo{person}{Max~L. Wilson}} (Eds.).
  \bibinfo{publisher}{{ACM}}, \bibinfo{pages}{365:1--365:22}.
\newblock
\urldef\tempurl%
\url{https://doi.org/10.1145/3544548.3581148}
\showDOI{\tempurl}


\bibitem[\protect\citeauthoryear{Hunter, Maes, Tang, Inkpen, and Hessey}{Hunter
  et~al\mbox{.}}{2014}]%
        {Hunter-CHI14}
\bibfield{author}{\bibinfo{person}{Seth~E. Hunter}, \bibinfo{person}{Pattie
  Maes}, \bibinfo{person}{Anthony Tang}, \bibinfo{person}{Kori~M. Inkpen},
  {and} \bibinfo{person}{Susan~M. Hessey}.} \bibinfo{year}{2014}\natexlab{}.
\newblock \showarticletitle{WaaZam!: supporting creative play at a distance in
  customized video environments}. In \bibinfo{booktitle}{\emph{{CHI} Conference
  on Human Factors in Computing Systems, CHI'14, Toronto, ON, Canada - April 26
  - May 01, 2014}}, \bibfield{editor}{\bibinfo{person}{Matt Jones},
  \bibinfo{person}{Philippe~A. Palanque}, \bibinfo{person}{Albrecht Schmidt},
  {and} \bibinfo{person}{Tovi Grossman}} (Eds.). \bibinfo{publisher}{{ACM}},
  \bibinfo{pages}{1197--1206}.
\newblock
\urldef\tempurl%
\url{https://doi.org/10.1145/2556288.2557382}
\showDOI{\tempurl}


\bibitem[\protect\citeauthoryear{Hyrkas, Wilson, Tang, Gamper, Sodoma,
  Tankelevitch, Inkpen, Chappidi, and Jones}{Hyrkas et~al\mbox{.}}{2023}]%
        {Hyrkas-CHI23}
\bibfield{author}{\bibinfo{person}{Jeremy Hyrkas}, \bibinfo{person}{Andrew~D.
  Wilson}, \bibinfo{person}{John Tang}, \bibinfo{person}{Hannes Gamper},
  \bibinfo{person}{Hong Sodoma}, \bibinfo{person}{Lev Tankelevitch},
  \bibinfo{person}{Kori Inkpen}, \bibinfo{person}{Shreya Chappidi}, {and}
  \bibinfo{person}{Brennan Jones}.} \bibinfo{year}{2023}\natexlab{}.
\newblock \showarticletitle{Spatialized Audio and Hybrid Video Conferencing:
  Where Should Voices be Positioned for People in the Room and Remote Headset
  Users?}. In \bibinfo{booktitle}{\emph{Proceedings of the 2023 {CHI}
  Conference on Human Factors in Computing Systems, {CHI} 2023, Hamburg,
  Germany, April 23-28, 2023}}, \bibfield{editor}{\bibinfo{person}{Albrecht
  Schmidt}, \bibinfo{person}{Kaisa V{\"{a}}{\"{a}}n{\"{a}}nen},
  \bibinfo{person}{Tesh Goyal}, \bibinfo{person}{Per~Ola Kristensson},
  \bibinfo{person}{Anicia Peters}, \bibinfo{person}{Stefanie Mueller},
  \bibinfo{person}{Julie~R. Williamson}, {and} \bibinfo{person}{Max~L. Wilson}}
  (Eds.). \bibinfo{publisher}{{ACM}}, \bibinfo{pages}{794:1--794:14}.
\newblock
\urldef\tempurl%
\url{https://doi.org/10.1145/3544548.3581085}
\showDOI{\tempurl}


\bibitem[\protect\citeauthoryear{Jr.}{Jr.}{2007}]%
        {Olsen-UIST07}
\bibfield{author}{\bibinfo{person}{Dan R.~Olsen Jr.}}
  \bibinfo{year}{2007}\natexlab{}.
\newblock \showarticletitle{Evaluating user interface systems research}. In
  \bibinfo{booktitle}{\emph{Proceedings of the 20th Annual {ACM} Symposium on
  User Interface Software and Technology, Newport, Rhode Island, USA, October
  7-10, 2007}}. \bibinfo{publisher}{{ACM}}, \bibinfo{pages}{251--258}.
\newblock
\urldef\tempurl%
\url{https://doi.org/10.1145/1294211.1294256}
\showDOI{\tempurl}


\bibitem[\protect\citeauthoryear{Jun and Nichol}{Jun and Nichol}{2023}]%
        {junShapEGeneratingConditional2023}
\bibfield{author}{\bibinfo{person}{Heewoo Jun} {and} \bibinfo{person}{Alex
  Nichol}.} \bibinfo{year}{2023}\natexlab{}.
\newblock \bibinfo{title}{Shap-{{E}}: {{Generating Conditional 3D Implicit
  Functions}}}.
\newblock
\newblock
\urldef\tempurl%
\url{https://doi.org/10.48550/arXiv.2305.02463}
\showDOI{\tempurl}
\showeprint[arxiv]{2305.02463}~[cs]


\bibitem[\protect\citeauthoryear{Junuzovic, Inkpen, Blank, and Gupta}{Junuzovic
  et~al\mbox{.}}{2012}]%
        {Junuzovic-CHI12}
\bibfield{author}{\bibinfo{person}{Sasa Junuzovic}, \bibinfo{person}{Kori
  Inkpen}, \bibinfo{person}{Tom Blank}, {and} \bibinfo{person}{Anoop Gupta}.}
  \bibinfo{year}{2012}\natexlab{}.
\newblock \showarticletitle{IllumiShare: sharing any surface}. In
  \bibinfo{booktitle}{\emph{{CHI} Conference on Human Factors in Computing
  Systems, {CHI} '12, Austin, TX, {USA} - May 05 - 10, 2012}},
  \bibfield{editor}{\bibinfo{person}{Joseph~A. Konstan}, \bibinfo{person}{Ed~H.
  Chi}, {and} \bibinfo{person}{Kristina H{\"{o}}{\"{o}}k}} (Eds.).
  \bibinfo{publisher}{{ACM}}, \bibinfo{pages}{1919--1928}.
\newblock
\urldef\tempurl%
\url{https://doi.org/10.1145/2207676.2208333}
\showDOI{\tempurl}


\bibitem[\protect\citeauthoryear{Karnewar, Vedaldi, Novotny, and
  Mitra}{Karnewar et~al\mbox{.}}{2023}]%
        {karnewarHOLODIFFUSIONTraining3D2023}
\bibfield{author}{\bibinfo{person}{Animesh Karnewar}, \bibinfo{person}{Andrea
  Vedaldi}, \bibinfo{person}{David Novotny}, {and} \bibinfo{person}{Niloy~J.
  Mitra}.} \bibinfo{year}{2023}\natexlab{}.
\newblock \showarticletitle{{{HOLODIFFUSION}}: {{Training}} a {{3D Diffusion
  Model Using 2D Images}}}. In \bibinfo{booktitle}{\emph{2023 {{IEEE}}/{{CVF
  Conference}} on {{Computer Vision}} and {{Pattern Recognition}} ({{CVPR}})}}.
  \bibinfo{publisher}{{IEEE}}, \bibinfo{address}{{Vancouver, BC, Canada}},
  \bibinfo{pages}{18423--18433}.
\newblock
\showISBNx{9798350301298}
\urldef\tempurl%
\url{https://doi.org/10.1109/CVPR52729.2023.01767}
\showDOI{\tempurl}


\bibitem[\protect\citeauthoryear{Koh, Agrawal, Batra, Tucker, Waters, Lee,
  Yang, Baldridge, and Anderson}{Koh et~al\mbox{.}}{2022}]%
        {kohSimpleEffectiveSynthesis2022}
\bibfield{author}{\bibinfo{person}{Jing~Yu Koh}, \bibinfo{person}{Harsh
  Agrawal}, \bibinfo{person}{Dhruv Batra}, \bibinfo{person}{Richard Tucker},
  \bibinfo{person}{Austin Waters}, \bibinfo{person}{Honglak Lee},
  \bibinfo{person}{Yinfei Yang}, \bibinfo{person}{Jason Baldridge}, {and}
  \bibinfo{person}{Peter Anderson}.} \bibinfo{year}{2022}\natexlab{}.
\newblock \bibinfo{title}{Simple and {{Effective Synthesis}} of {{Indoor 3D
  Scenes}}}.
\newblock
\newblock
\urldef\tempurl%
\url{https://doi.org/10.48550/arXiv.2204.02960}
\showDOI{\tempurl}
\showeprint[arxiv]{2204.02960}~[cs]


\bibitem[\protect\citeauthoryear{Kumaravel, Anderson, Fitzmaurice, Hartmann,
  and Grossman}{Kumaravel et~al\mbox{.}}{2019}]%
        {Kumaravel-UIST19}
\bibfield{author}{\bibinfo{person}{Balasaravanan~Thoravi Kumaravel},
  \bibinfo{person}{Fraser Anderson}, \bibinfo{person}{George~W. Fitzmaurice},
  \bibinfo{person}{Bjoern Hartmann}, {and} \bibinfo{person}{Tovi Grossman}.}
  \bibinfo{year}{2019}\natexlab{}.
\newblock \showarticletitle{Loki: Facilitating Remote Instruction of Physical
  Tasks Using Bi-Directional Mixed-Reality Telepresence}. In
  \bibinfo{booktitle}{\emph{Proceedings of the 32nd Annual {ACM} Symposium on
  User Interface Software and Technology, {UIST} 2019, New Orleans, LA, USA,
  October 20-23, 2019}}, \bibfield{editor}{\bibinfo{person}{Fran{\c{c}}ois
  Guimbreti{\`{e}}re}, \bibinfo{person}{Michael~S. Bernstein}, {and}
  \bibinfo{person}{Katharina Reinecke}} (Eds.). \bibinfo{publisher}{{ACM}},
  \bibinfo{pages}{161--174}.
\newblock
\urldef\tempurl%
\url{https://doi.org/10.1145/3332165.3347872}
\showDOI{\tempurl}


\bibitem[\protect\citeauthoryear{Labrie, Mok, Tang, Lui, Oehlberg, and
  Poretski}{Labrie et~al\mbox{.}}{2022}]%
        {Labrie-GROUP22}
\bibfield{author}{\bibinfo{person}{Audrey Labrie}, \bibinfo{person}{Terrance
  Tin~Hoi Mok}, \bibinfo{person}{Anthony Tang}, \bibinfo{person}{Michelle Lui},
  \bibinfo{person}{Lora Oehlberg}, {and} \bibinfo{person}{Lev Poretski}.}
  \bibinfo{year}{2022}\natexlab{}.
\newblock \showarticletitle{Toward Video-Conferencing Tools for Hands-On
  Activities in Online Teaching}.
\newblock \bibinfo{journal}{\emph{Proc. {ACM} Hum. Comput. Interact.}}
  \bibinfo{volume}{6}, \bibinfo{number}{{GROUP}} (\bibinfo{year}{2022}),
  \bibinfo{pages}{10:1--10:22}.
\newblock
\urldef\tempurl%
\url{https://doi.org/10.1145/3492829}
\showDOI{\tempurl}


\bibitem[\protect\citeauthoryear{Lee, Park, Lee, and Lee}{Lee
  et~al\mbox{.}}{2022}]%
        {Lee-CHI22}
\bibfield{author}{\bibinfo{person}{Minha Lee}, \bibinfo{person}{Wonyoung Park},
  \bibinfo{person}{Sunok Lee}, {and} \bibinfo{person}{Sangsu Lee}.}
  \bibinfo{year}{2022}\natexlab{}.
\newblock \showarticletitle{Distracting Moments in Videoconferencing: {A} Look
  Back at the Pandemic Period}. In \bibinfo{booktitle}{\emph{{CHI} '22: {CHI}
  Conference on Human Factors in Computing Systems, New Orleans, LA, USA, 29
  April 2022 - 5 May 2022}}, \bibfield{editor}{\bibinfo{person}{Simone D.~J.
  Barbosa}, \bibinfo{person}{Cliff Lampe}, \bibinfo{person}{Caroline Appert},
  \bibinfo{person}{David~A. Shamma}, \bibinfo{person}{Steven~Mark Drucker},
  \bibinfo{person}{Julie~R. Williamson}, {and} \bibinfo{person}{Koji Yatani}}
  (Eds.). \bibinfo{publisher}{{ACM}}, \bibinfo{pages}{141:1--141:21}.
\newblock
\urldef\tempurl%
\url{https://doi.org/10.1145/3491102.3517545}
\showDOI{\tempurl}


\bibitem[\protect\citeauthoryear{Li and Bansal}{Li and Bansal}{2023}]%
        {li2023panogen}
\bibfield{author}{\bibinfo{person}{Jialu Li} {and} \bibinfo{person}{Mohit
  Bansal}.} \bibinfo{year}{2023}\natexlab{}.
\newblock \showarticletitle{PanoGen: Text-Conditioned Panoramic Environment
  Generation for Vision-and-Language Navigation}.
\newblock \bibinfo{journal}{\emph{arxiv}} (\bibinfo{year}{2023}).
\newblock


\bibitem[\protect\citeauthoryear{Li, Liu, Wu, Mu, Yang, Gao, Li, and Lee}{Li
  et~al\mbox{.}}{2023}]%
        {liGligenOpensetGrounded2023}
\bibfield{author}{\bibinfo{person}{Yuheng Li}, \bibinfo{person}{Haotian Liu},
  \bibinfo{person}{Qingyang Wu}, \bibinfo{person}{Fangzhou Mu},
  \bibinfo{person}{Jianwei Yang}, \bibinfo{person}{Jianfeng Gao},
  \bibinfo{person}{Chunyuan Li}, {and} \bibinfo{person}{Yong~Jae Lee}.}
  \bibinfo{year}{2023}\natexlab{}.
\newblock \showarticletitle{{{GLIGEN}}: {{Open-Set Grounded Text-to-Image
  Generation}}}. In \bibinfo{booktitle}{\emph{Proceedings of the {{IEEE}}/{{CVF
  Conference}} on {{Computer Vision}} and {{Pattern Recognition}}}}.
  \bibinfo{pages}{22511--22521}.
\newblock


\bibitem[\protect\citeauthoryear{Liu, Qiao, and Chilton}{Liu
  et~al\mbox{.}}{2022}]%
        {liuOpal2022}
\bibfield{author}{\bibinfo{person}{Vivian Liu}, \bibinfo{person}{Han Qiao},
  {and} \bibinfo{person}{Lydia~B. Chilton}.} \bibinfo{year}{2022}\natexlab{}.
\newblock \showarticletitle{Opal: Multimodal Image Generation for News
  Illustration}. In \bibinfo{booktitle}{\emph{The 35th Annual {ACM} Symposium
  on User Interface Software and Technology, {UIST} 2022, Bend, OR, USA, 29
  October 2022 - 2 November 2022}}. \bibinfo{publisher}{{ACM}},
  \bibinfo{pages}{73:1--73:17}.
\newblock
\urldef\tempurl%
\url{https://doi.org/10.1145/3526113.3545621}
\showDOI{\tempurl}


\bibitem[\protect\citeauthoryear{Mirza and Osindero}{Mirza and
  Osindero}{2014}]%
        {mirzaConditionalGenerativeAdversarial2014}
\bibfield{author}{\bibinfo{person}{Mehdi Mirza} {and} \bibinfo{person}{Simon
  Osindero}.} \bibinfo{year}{2014}\natexlab{}.
\newblock \bibinfo{title}{Conditional {{Generative Adversarial Nets}}}.
\newblock
\newblock
\urldef\tempurl%
\url{https://doi.org/10.48550/arXiv.1411.1784}
\showDOI{\tempurl}
\showeprint[arxiv]{1411.1784}~[cs, stat]


\bibitem[\protect\citeauthoryear{Morikawa and Maesako}{Morikawa and
  Maesako}{1998}]%
        {Morikawa-CHI98}
\bibfield{author}{\bibinfo{person}{Osamu Morikawa} {and}
  \bibinfo{person}{Takanori Maesako}.} \bibinfo{year}{1998}\natexlab{}.
\newblock \showarticletitle{HyperMirror: Toward Pleasant-to-Use Video Mediated
  Communication System}. In \bibinfo{booktitle}{\emph{{CSCW} '98, Proceedings
  of the {ACM} 1998 Conference on Computer Supported Cooperative Work, Seattle,
  WA, USA, November 14-18, 1998}}, \bibfield{editor}{\bibinfo{person}{Steven~E.
  Poltrock} {and} \bibinfo{person}{Jonathan Grudin}} (Eds.).
  \bibinfo{publisher}{{ACM}}, \bibinfo{pages}{149--158}.
\newblock
\urldef\tempurl%
\url{https://doi.org/10.1145/289444.289489}
\showDOI{\tempurl}


\bibitem[\protect\citeauthoryear{Mu, Borowski, Gr{\o}nb{\ae}k, B{\o}dker, and
  Hoggan}{Mu et~al\mbox{.}}{2024}]%
        {Mu-WhisperThroughWalls-CHI24}
\bibfield{author}{\bibinfo{person}{Qianqian Mu}, \bibinfo{person}{Marcel
  Borowski}, \bibinfo{person}{Jens Emil~Sloth Gr{\o}nb{\ae}k},
  \bibinfo{person}{Susanne B{\o}dker}, {and} \bibinfo{person}{Eve~E. Hoggan}.}
  \bibinfo{year}{2024}\natexlab{}.
\newblock \showarticletitle{Whispering Through Walls: Towards Inclusive
  Backchannel Communication in Hybrid Meetings}. In
  \bibinfo{booktitle}{\emph{Proceedings of the {CHI} Conference on Human
  Factors in Computing Systems, {CHI} 2024, Honolulu, HI, USA, May 11-16,
  2024}}. \bibinfo{publisher}{{ACM}}, \bibinfo{pages}{1032:1--1032:16}.
\newblock
\urldef\tempurl%
\url{https://doi.org/10.1145/3613904.3642419}
\showDOI{\tempurl}


\bibitem[\protect\citeauthoryear{Numan, Giunchi, Congdon, and Steed}{Numan
  et~al\mbox{.}}{2023}]%
        {numanUbiqGenieLeveragingExternal2023}
\bibfield{author}{\bibinfo{person}{Nels Numan}, \bibinfo{person}{Daniele
  Giunchi}, \bibinfo{person}{Benjamin Congdon}, {and} \bibinfo{person}{Anthony
  Steed}.} \bibinfo{year}{2023}\natexlab{}.
\newblock \showarticletitle{Ubiq-{{Genie}}: {{Leveraging External Frameworks}}
  for {{Enhanced Social VR Experiences}}}. In \bibinfo{booktitle}{\emph{2023
  {{IEEE Conference}} on {{Virtual Reality}} and {{3D User Interfaces
  Abstracts}} and {{Workshops}} ({{VRW}})}}. \bibinfo{publisher}{{IEEE}},
  \bibinfo{address}{{Shanghai, China}}, \bibinfo{pages}{497--501}.
\newblock
\urldef\tempurl%
\url{https://doi.org/10.1109/VRW58643.2023.00108}
\showDOI{\tempurl}


\bibitem[\protect\citeauthoryear{Numan, Rajaram, Kumaravel, Marquardt, and
  Wilson}{Numan et~al\mbox{.}}{2024}]%
        {Numan-UIST24}
\bibfield{author}{\bibinfo{person}{Nels Numan}, \bibinfo{person}{Shwetha
  Rajaram}, \bibinfo{person}{Balasaravanan~Thoravi Kumaravel},
  \bibinfo{person}{Nicolai Marquardt}, {and} \bibinfo{person}{Andrew~D.
  Wilson}.} \bibinfo{year}{2024}\natexlab{}.
\newblock \showarticletitle{SpaceBlender: Creating Context-Rich Collaborative
  Spaces Through Generative 3D Scene Blending}. In
  \bibinfo{booktitle}{\emph{Proceedings of the 37th Annual {ACM} Symposium on
  User Interface Software and Technology, {UIST} 2024, Pittsburgh, PA, USA,
  October 13-16, 2024}}. \bibinfo{publisher}{{ACM}}.
\newblock
\urldef\tempurl%
\url{https://doi.org/10.1145/3654777.3676361}
\showDOI{\tempurl}


\bibitem[\protect\citeauthoryear{O'Hara, Kjeldskov, and Paay}{O'Hara
  et~al\mbox{.}}{2011}]%
        {OHara-TOCHI11}
\bibfield{author}{\bibinfo{person}{Kenton O'Hara}, \bibinfo{person}{Jesper
  Kjeldskov}, {and} \bibinfo{person}{Jeni Paay}.}
  \bibinfo{year}{2011}\natexlab{}.
\newblock \showarticletitle{Blended interaction spaces for distributed team
  collaboration}.
\newblock \bibinfo{journal}{\emph{{ACM} Trans. Comput. Hum. Interact.}}
  \bibinfo{volume}{18}, \bibinfo{number}{1} (\bibinfo{year}{2011}),
  \bibinfo{pages}{3:1--3:28}.
\newblock
\urldef\tempurl%
\url{https://doi.org/10.1145/1959022.1959025}
\showDOI{\tempurl}


\bibitem[\protect\citeauthoryear{Olafenwa}{Olafenwa}{2021}]%
        {olafenwaSimplifyingObjectSegmentation2021}
\bibfield{author}{\bibinfo{person}{Ayoola Olafenwa}.}
  \bibinfo{year}{2021}\natexlab{}.
\newblock \showarticletitle{Simplifying {{Object Segmentation}} with {{PixelLib
  Library}}}.
\newblock  (\bibinfo{date}{Jan.} \bibinfo{year}{2021}).
\newblock


\bibitem[\protect\citeauthoryear{OpenAI}{OpenAI}{2023}]%
        {openaiGPT4TechnicalReport2023}
\bibfield{author}{\bibinfo{person}{OpenAI}.} \bibinfo{year}{2023}\natexlab{}.
\newblock \bibinfo{title}{{{GPT-4 Technical Report}}}.
\newblock
\newblock
\urldef\tempurl%
\url{https://doi.org/10.48550/arXiv.2303.08774}
\showDOI{\tempurl}
\showeprint[arxiv]{2303.08774}~[cs]


\bibitem[\protect\citeauthoryear{Pejsa, Kantor, Benko, Ofek, and Wilson}{Pejsa
  et~al\mbox{.}}{2016}]%
        {Pejsa-CSCW16}
\bibfield{author}{\bibinfo{person}{Tomislav Pejsa}, \bibinfo{person}{Julian
  Kantor}, \bibinfo{person}{Hrvoje Benko}, \bibinfo{person}{Eyal Ofek}, {and}
  \bibinfo{person}{Andrew~D. Wilson}.} \bibinfo{year}{2016}\natexlab{}.
\newblock \showarticletitle{Room2Room: Enabling Life-Size Telepresence in a
  Projected Augmented Reality Environment}. In
  \bibinfo{booktitle}{\emph{Proceedings of the 19th {ACM} Conference on
  Computer-Supported Cooperative Work {\&} Social Computing, {CSCW} 2016, San
  Francisco, CA, USA, February 27 - March 2, 2016}},
  \bibfield{editor}{\bibinfo{person}{Darren Gergle},
  \bibinfo{person}{Meredith~Ringel Morris}, \bibinfo{person}{Pernille
  Bj{\o}rn}, {and} \bibinfo{person}{Joseph~A. Konstan}} (Eds.).
  \bibinfo{publisher}{{ACM}}, \bibinfo{pages}{1714--1723}.
\newblock
\urldef\tempurl%
\url{https://doi.org/10.1145/2818048.2819965}
\showDOI{\tempurl}


\bibitem[\protect\citeauthoryear{Radford, Narasimhan, Salimans, and
  Sutskever}{Radford et~al\mbox{.}}{2018}]%
        {radfordImprovingLanguageUnderstanding2018}
\bibfield{author}{\bibinfo{person}{Alec Radford}, \bibinfo{person}{Karthik
  Narasimhan}, \bibinfo{person}{Tim Salimans}, {and} \bibinfo{person}{Ilya
  Sutskever}.} \bibinfo{year}{2018}\natexlab{}.
\newblock \bibinfo{title}{Improving {{Language Understanding}} by {{Generative
  Pre-Training}}}.
\newblock
\newblock


\bibitem[\protect\citeauthoryear{Rombach, Blattmann, Lorenz, Esser, and
  Ommer}{Rombach et~al\mbox{.}}{2022}]%
        {rombachHighResolutionImageSynthesis2022}
\bibfield{author}{\bibinfo{person}{Robin Rombach}, \bibinfo{person}{Andreas
  Blattmann}, \bibinfo{person}{Dominik Lorenz}, \bibinfo{person}{Patrick
  Esser}, {and} \bibinfo{person}{Bjorn Ommer}.}
  \bibinfo{year}{2022}\natexlab{}.
\newblock \showarticletitle{High-{{Resolution Image Synthesis}} with {{Latent
  Diffusion Models}}}. In \bibinfo{booktitle}{\emph{2022 {{IEEE}}/{{CVF
  Conference}} on {{Computer Vision}} and {{Pattern Recognition}} ({{CVPR}})}}.
  \bibinfo{publisher}{{IEEE}}, \bibinfo{address}{{New Orleans, LA, USA}},
  \bibinfo{pages}{10674--10685}.
\newblock
\showISBNx{978-1-66546-946-3}
\urldef\tempurl%
\url{https://doi.org/10.1109/CVPR52688.2022.01042}
\showDOI{\tempurl}


\bibitem[\protect\citeauthoryear{Scupin}{Scupin}{1997}]%
        {Scupin-1997}
\bibfield{author}{\bibinfo{person}{Raymond Scupin}.}
  \bibinfo{year}{1997}\natexlab{}.
\newblock \showarticletitle{The {{KJ Method}}: {{A Technique}} for {{Analyzing
  Data Derived}} from {{Japanese Ethnology}}}.
\newblock \bibinfo{journal}{\emph{Human Organization}} \bibinfo{volume}{56},
  \bibinfo{number}{2} (\bibinfo{year}{1997}), \bibinfo{pages}{233--237}.
\newblock
\showISSN{0018-7259}


\bibitem[\protect\citeauthoryear{Sharma, Suhubdy, Michalski, Kahou, and
  Bengio}{Sharma et~al\mbox{.}}{2018}]%
        {sharmaChatPainterImprovingText2018}
\bibfield{author}{\bibinfo{person}{Shikhar Sharma}, \bibinfo{person}{Dendi
  Suhubdy}, \bibinfo{person}{Vincent Michalski},
  \bibinfo{person}{Samira~Ebrahimi Kahou}, {and} \bibinfo{person}{Yoshua
  Bengio}.} \bibinfo{year}{2018}\natexlab{}.
\newblock \bibinfo{title}{{{ChatPainter}}: {{Improving Text}} to {{Image
  Generation}} Using {{Dialogue}}}.
\newblock
\newblock
\urldef\tempurl%
\url{https://doi.org/10.48550/arXiv.1802.08216}
\showDOI{\tempurl}
\showeprint[arxiv]{1802.08216}~[cs]


\bibitem[\protect\citeauthoryear{Tang, Inkpen, Junuzovic, Mallari, Rintel,
  Wilson, Cupala, Carbary, Sellen, and Buxton}{Tang et~al\mbox{.}}{2023}]%
        {Tang-CSCW23}
\bibfield{author}{\bibinfo{person}{John Tang}, \bibinfo{person}{Kori Inkpen},
  \bibinfo{person}{Sasa Junuzovic}, \bibinfo{person}{Keri Mallari},
  \bibinfo{person}{Sean Rintel}, \bibinfo{person}{Andrew Wilson},
  \bibinfo{person}{Shiraz Cupala}, \bibinfo{person}{Tony Carbary},
  \bibinfo{person}{Abigail Sellen}, {and} \bibinfo{person}{William Buxton}.}
  \bibinfo{year}{2023}\natexlab{}.
\newblock \showarticletitle{Perspectives: {{Creating Inclusive}} and
  {{Equitable Hybrid Meeting Experiences}}}.
\newblock \bibinfo{journal}{\emph{Proceedings of the ACM on Human-Computer
  Interaction}} \bibinfo{volume}{7}, \bibinfo{number}{CSCW2}
  (\bibinfo{date}{Oct.} \bibinfo{year}{2023}).
\newblock


\bibitem[\protect\citeauthoryear{Tuddenham and Robinson}{Tuddenham and
  Robinson}{2009}]%
        {Tuddenham-CHI09}
\bibfield{author}{\bibinfo{person}{Philip Tuddenham} {and}
  \bibinfo{person}{Peter Robinson}.} \bibinfo{year}{2009}\natexlab{}.
\newblock \showarticletitle{Territorial coordination and workspace awareness in
  remote tabletop collaboration}. In \bibinfo{booktitle}{\emph{Proceedings of
  the 27th International Conference on Human Factors in Computing Systems,
  {CHI} 2009, Boston, MA, USA, April 4-9, 2009}}. \bibinfo{publisher}{{ACM}},
  \bibinfo{pages}{2139--2148}.
\newblock
\urldef\tempurl%
\url{https://doi.org/10.1145/1518701.1519026}
\showDOI{\tempurl}


\bibitem[\protect\citeauthoryear{Venolia, Tang, Inkpen, and Unver}{Venolia
  et~al\mbox{.}}{2018}]%
        {Venolia-MobileHCI18}
\bibfield{author}{\bibinfo{person}{Gina Venolia}, \bibinfo{person}{John~C.
  Tang}, \bibinfo{person}{Kori Inkpen}, {and} \bibinfo{person}{Baris Unver}.}
  \bibinfo{year}{2018}\natexlab{}.
\newblock \showarticletitle{Wish you were here: being together through
  composite video and digital keepsakes}. In
  \bibinfo{booktitle}{\emph{Proceedings of the 20th International Conference on
  Human-Computer Interaction with Mobile Devices and Services, MobileHCI 2018,
  Barcelona, Spain, September 03-06, 2018}},
  \bibfield{editor}{\bibinfo{person}{Lynne Baillie} {and}
  \bibinfo{person}{Nuria Oliver}} (Eds.). \bibinfo{publisher}{{ACM}},
  \bibinfo{pages}{17:1--17:11}.
\newblock
\urldef\tempurl%
\url{https://doi.org/10.1145/3229434.3229476}
\showDOI{\tempurl}


\bibitem[\protect\citeauthoryear{Xia, Herscher, Perlin, and Wigdor}{Xia
  et~al\mbox{.}}{2018}]%
        {Xia-Spacetime-UIST18}
\bibfield{author}{\bibinfo{person}{Haijun Xia}, \bibinfo{person}{Sebastian
  Herscher}, \bibinfo{person}{Ken Perlin}, {and} \bibinfo{person}{Daniel
  Wigdor}.} \bibinfo{year}{2018}\natexlab{}.
\newblock \showarticletitle{Spacetime: Enabling Fluid Individual and
  Collaborative Editing in Virtual Reality}. In \bibinfo{booktitle}{\emph{The
  31st Annual {ACM} Symposium on User Interface Software and Technology, {UIST}
  2018, Berlin, Germany, October 14-17, 2018}}. \bibinfo{publisher}{{ACM}},
  \bibinfo{pages}{853--866}.
\newblock
\urldef\tempurl%
\url{https://doi.org/10.1145/3242587.3242597}
\showDOI{\tempurl}


\bibitem[\protect\citeauthoryear{Yang, Holz, Ofek, and Wilson}{Yang
  et~al\mbox{.}}{2019}]%
        {yangDreamWalkerSubstitutingRealWorld2019}
\bibfield{author}{\bibinfo{person}{Jackie~(Junrui) Yang},
  \bibinfo{person}{Christian Holz}, \bibinfo{person}{Eyal Ofek}, {and}
  \bibinfo{person}{Andrew~D. Wilson}.} \bibinfo{year}{2019}\natexlab{}.
\newblock \showarticletitle{{{DreamWalker}}: {{Substituting Real-World Walking
  Experiences}} with a {{Virtual Reality}}}. In
  \bibinfo{booktitle}{\emph{Proceedings of the 32nd {{Annual ACM Symposium}} on
  {{User Interface Software}} and {{Technology}}}}
  \emph{(\bibinfo{series}{{{UIST}} '19})}. \bibinfo{publisher}{{Association for
  Computing Machinery}}, \bibinfo{address}{{New York, NY, USA}},
  \bibinfo{pages}{1093--1107}.
\newblock
\showISBNx{978-1-4503-6816-2}
\urldef\tempurl%
\url{https://doi.org/10.1145/3332165.3347875}
\showDOI{\tempurl}


\bibitem[\protect\citeauthoryear{Yu, Lin, Yang, Shen, Lu, and Huang}{Yu
  et~al\mbox{.}}{2019}]%
        {yuFreeFormImageInpainting2019}
\bibfield{author}{\bibinfo{person}{Jiahui Yu}, \bibinfo{person}{Zhe Lin},
  \bibinfo{person}{Jimei Yang}, \bibinfo{person}{Xiaohui Shen},
  \bibinfo{person}{Xin Lu}, {and} \bibinfo{person}{Thomas Huang}.}
  \bibinfo{year}{2019}\natexlab{}.
\newblock \showarticletitle{Free-{{Form Image Inpainting With Gated
  Convolution}}}. In \bibinfo{booktitle}{\emph{2019 {{IEEE}}/{{CVF
  International Conference}} on {{Computer Vision}} ({{ICCV}})}}.
  \bibinfo{publisher}{{IEEE}}, \bibinfo{address}{{Seoul, Korea (South)}},
  \bibinfo{pages}{4470--4479}.
\newblock
\showISBNx{978-1-72814-803-8}
\urldef\tempurl%
\url{https://doi.org/10.1109/ICCV.2019.00457}
\showDOI{\tempurl}


\bibitem[\protect\citeauthoryear{Yuan, Cao, Wang, and Yarosh}{Yuan
  et~al\mbox{.}}{2021}]%
        {Yuan-CHI21}
\bibfield{author}{\bibinfo{person}{Ye Yuan}, \bibinfo{person}{Jan Cao},
  \bibinfo{person}{Ruotong Wang}, {and} \bibinfo{person}{Svetlana Yarosh}.}
  \bibinfo{year}{2021}\natexlab{}.
\newblock \showarticletitle{Tabletop Games in the Age of Remote Collaboration:
  Design Opportunities for a Socially Connected Game Experience}. In
  \bibinfo{booktitle}{\emph{{CHI} '21: {CHI} Conference on Human Factors in
  Computing Systems, Virtual Event / Yokohama, Japan, May 8-13, 2021}},
  \bibfield{editor}{\bibinfo{person}{Yoshifumi Kitamura},
  \bibinfo{person}{Aaron Quigley}, \bibinfo{person}{Katherine Isbister},
  \bibinfo{person}{Takeo Igarashi}, \bibinfo{person}{Pernille Bj{\o}rn}, {and}
  \bibinfo{person}{Steven~Mark Drucker}} (Eds.). \bibinfo{publisher}{{ACM}},
  \bibinfo{pages}{436:1--436:14}.
\newblock
\urldef\tempurl%
\url{https://doi.org/10.1145/3411764.3445512}
\showDOI{\tempurl}


\bibitem[\protect\citeauthoryear{Zagermann, Pfeil, R{\"{a}}dle, Jetter,
  Klokmose, and Reiterer}{Zagermann et~al\mbox{.}}{2016}]%
        {Zagermann-CHI16}
\bibfield{author}{\bibinfo{person}{Johannes Zagermann}, \bibinfo{person}{Ulrike
  Pfeil}, \bibinfo{person}{Roman R{\"{a}}dle},
  \bibinfo{person}{Hans{-}Christian Jetter},
  \bibinfo{person}{Clemens~Nylandsted Klokmose}, {and} \bibinfo{person}{Harald
  Reiterer}.} \bibinfo{year}{2016}\natexlab{}.
\newblock \showarticletitle{When Tablets meet Tabletops: The Effect of Tabletop
  Size on Around-the-Table Collaboration with Personal Tablets}. In
  \bibinfo{booktitle}{\emph{Proceedings of the 2016 {CHI} Conference on Human
  Factors in Computing Systems, San Jose, CA, USA, May 7-12, 2016}}.
  \bibinfo{publisher}{{ACM}}, \bibinfo{pages}{5470--5481}.
\newblock
\urldef\tempurl%
\url{https://doi.org/10.1145/2858036.2858224}
\showDOI{\tempurl}


\bibitem[\protect\citeauthoryear{Zhang, Agrawal, Oney, and Guo}{Zhang
  et~al\mbox{.}}{2023a}]%
        {Zhang-VRGit-CHI23}
\bibfield{author}{\bibinfo{person}{Lei Zhang}, \bibinfo{person}{Ashutosh
  Agrawal}, \bibinfo{person}{Steve Oney}, {and} \bibinfo{person}{Anhong Guo}.}
  \bibinfo{year}{2023}\natexlab{a}.
\newblock \showarticletitle{VRGit: {A} Version Control System for Collaborative
  Content Creation in Virtual Reality}. In
  \bibinfo{booktitle}{\emph{Proceedings of the 2023 {CHI} Conference on Human
  Factors in Computing Systems, {CHI} 2023, Hamburg, Germany, April 23-28,
  2023}}. \bibinfo{publisher}{{ACM}}, \bibinfo{pages}{36:1--36:14}.
\newblock
\urldef\tempurl%
\url{https://doi.org/10.1145/3544548.3581136}
\showDOI{\tempurl}


\bibitem[\protect\citeauthoryear{Zhang, Rao, and Agrawala}{Zhang
  et~al\mbox{.}}{2023b}]%
        {zhangAddingConditionalControl2023}
\bibfield{author}{\bibinfo{person}{Lvmin Zhang}, \bibinfo{person}{Anyi Rao},
  {and} \bibinfo{person}{Maneesh Agrawala}.} \bibinfo{year}{2023}\natexlab{b}.
\newblock \bibinfo{title}{Adding {{Conditional Control}} to {{Text-to-Image
  Diffusion Models}}}.
\newblock
\newblock
\urldef\tempurl%
\url{https://doi.org/10.48550/arXiv.2302.05543}
\showDOI{\tempurl}
\showeprint[arxiv]{2302.05543}~[cs]


\bibitem[\protect\citeauthoryear{Zhang, Zhu, Isola, Geng, Lin, Yu, and
  Efros}{Zhang et~al\mbox{.}}{2017}]%
        {zhangRealtimeUserguidedImage2017}
\bibfield{author}{\bibinfo{person}{Richard Zhang}, \bibinfo{person}{Jun-Yan
  Zhu}, \bibinfo{person}{Phillip Isola}, \bibinfo{person}{Xinyang Geng},
  \bibinfo{person}{Angela~S. Lin}, \bibinfo{person}{Tianhe Yu}, {and}
  \bibinfo{person}{Alexei~A. Efros}.} \bibinfo{year}{2017}\natexlab{}.
\newblock \showarticletitle{Real-Time User-Guided Image Colorization with
  Learned Deep Priors}.
\newblock \bibinfo{journal}{\emph{ACM Transactions on Graphics}}
  \bibinfo{volume}{36}, \bibinfo{number}{4} (\bibinfo{date}{July}
  \bibinfo{year}{2017}), \bibinfo{pages}{119:1--119:11}.
\newblock
\showISSN{0730-0301}
\urldef\tempurl%
\url{https://doi.org/10.1145/3072959.3073703}
\showDOI{\tempurl}


\end{thebibliography}

\onecolumn
\appendix
\section{Appendix}
\label{sec:appendix}

\subsection{Additional Examples of Blended Environments}
\label{appendix:additional-examples}
\begin{figure*}[htb!]
    \centering
    \includegraphics[width=\linewidth]{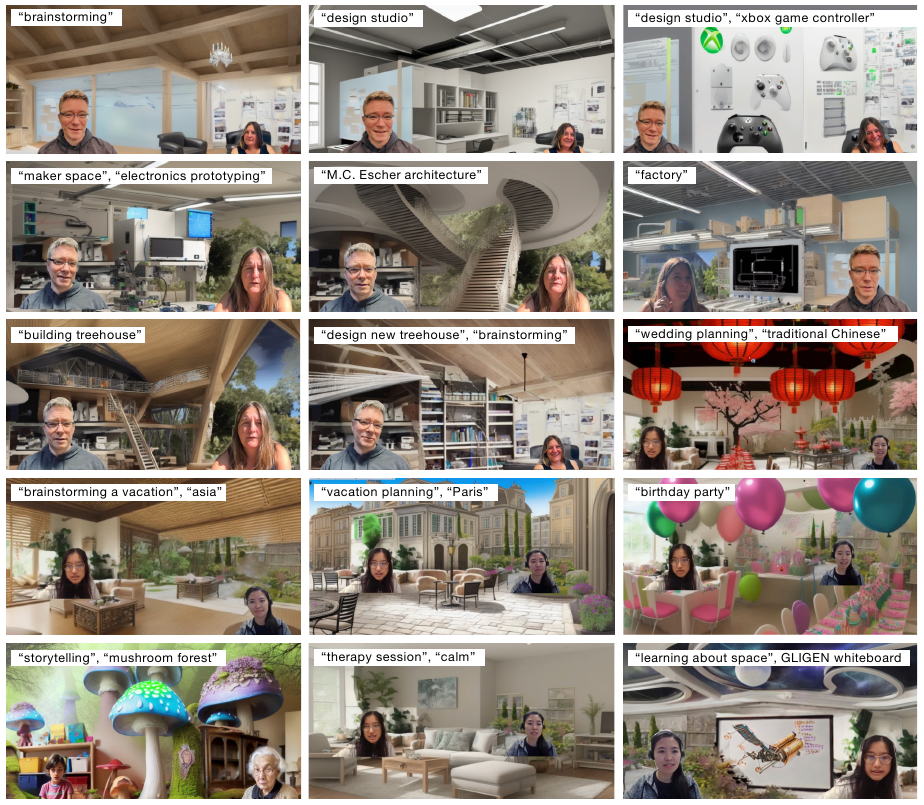}
    \caption{Examples of other generated meeting environments.}
    \Description{Overview of 12 examples of generated meeting environments, featuring prompts such as "factory", "building treehouse", "birthday party" and "wedding planning; traditional Chinese".}
    \label{fig:examples}
\end{figure*}

\newpage
\subsection{GPT-aided Prompt Crafting for Image Generation}
\label{appendix:GPT-aided-prompting}
\newcommand{\baseprompt}{\textcolor{green!60!black}{\texttt{Mushroom forest}-themed environment for a \texttt{brainstorming session}}}

Crafting textual prompts that result in desired output with image generation models (e.g., Stable Diffusion) requires significant expertise, and trial and error \cite{bradePromptifyTexttoImageGeneration2023}. To ease the process of prompt creation for stable diffusion, \SystemName enhances users' prompts by leveraging an LLM (in this case, GPT-3.5) to augment users' prompts with additional keywords related to the \textit{Meeting Activity} and \textit{Meeting Theme}. 
The advantage of this approach is that it allows for the dynamic generation of keywords for image generation that suit the varying needs of users in the meeting.

\paragraph{\textnormal{
For example, if the meeting activity is ''\texttt{brainstorming}'', and the theme of the meeting is ''\texttt{mushroom forest}'', then the system crafts a base prompt of \baseprompt.}} 
We then leverage single-shot prompting with GPT-3.5 to enhance the base prompt with contextually-relevant details:

\paragraph{\textnormal{\texttt{
\textbf{System Prompt to GPT}: Your task is to help the user create a Stable Diffusion prompt to generate an environment design.}}}
\texttt{
The user will specify an activity to occur in the environment and/or a theme for the space. You will provide a list of 4-5 types of objects to put in the environment and 4-5 distinct characteristics that describe the environment. The characteristics must be detailed and designed to generate visually appealing and cohesive results. 
Here is an example for a brainstorming activity: 
}

\noindent
\texttt{\{}

\paragraph{\textnormal{\texttt{Objects: ``whiteboards, plants, chairs, small tables''}}}

\paragraph{\textnormal{\texttt{Environment Characteristics:}}}
\texttt{``bright, open space, natural light, refreshing atmosphere, varied textures''}

\noindent
\texttt{\}}

\paragraph{\textnormal{Subsequently, depending on the base prompt derived from the user's input prompt, \SystemName crafts an input to GPT-3.5:}}

\paragraph{\textnormal{\texttt{
\textbf{Input Prompt (constructed by \SystemName) to GPT-3.5}: Provide a list of 4-5 types of objects to put in this environment and 4-5 characteristics that describe this environment: \baseprompt. Return the output as comma-separated strings in JSON format: \{Objects: string, Environment Characteristics: string\}.
}
}}

\paragraph{\textnormal{The results of such a prompting strategy to GPT results}} in the following output.

\noindent
\texttt{\{}

\paragraph{\textnormal{\texttt{Objects: ``Mushrooms, Trees, Fairies, Moss-covered rocks, Lanterns''}}}

\paragraph{\textnormal{\texttt{Environment Characteristics:}}}
\texttt{``Enchanting, Magical, Misty, Whimsical, Serene''}

\noindent
\texttt{\}}

\paragraph{\textnormal{
Additionally, \SystemName adds a fixed set of terms - ``\texttt{highly detailed, intricate, sharp focus, smooth}'' that we found to improve the results. 
In this case, the final prompt provided to Stable Diffusion is: }}

\paragraph{\textnormal{\texttt{\baseprompt; Giant mushrooms, Fairy houses, Moss-covered rocks, Glowing mushrooms, Enchanted flowers; Enchanting, Magical, Misty, Whimsical, Serene; highly detailed, intricate, sharp focus, smooth}}}

\paragraph{\textnormal{Fig.~\ref{fig:scenario2} from our \textit{Storytelling with Family} scenario shows the result of an image generation with such a prompt.}}

\newpage
\subsection{User Study Scenarios}
\label{appendix:user-study-scenarios}

\begin{figure*}[h!]
    \centering
    \includegraphics[width=\linewidth]{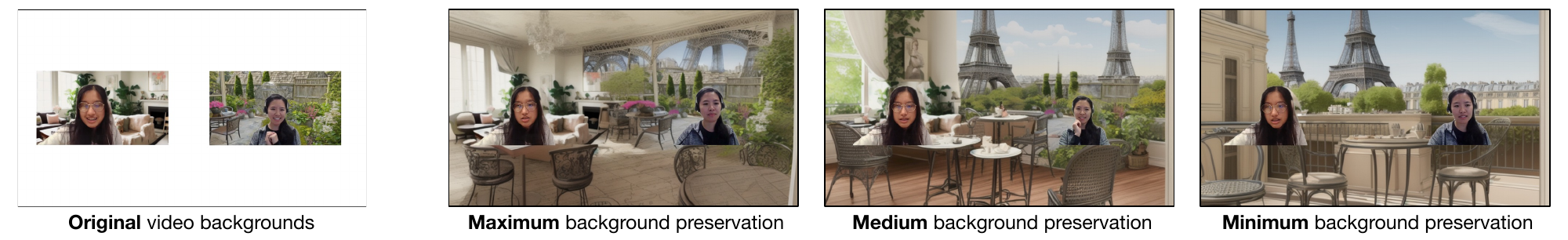}
    \caption{Vacation Planning Scenario: Two friends use \SystemName to get into the spirit of planning a trip to Paris. In Task 1 of our user study, we asked participants to compare three pre-canned environments that preserve their video backgrounds to different extents.
    }
    \label{fig:task1-envs}
    \Description{Overview of the "Vacation Planning Scenario". Two friends use BlendScape to get into the spirit of planning a trip to Paris. In Task 1 of our user study, we asked participants to compare three pre-canned environments that preserve their video backgrounds to different extents. This figure shows the three generated environments, with the first environment preserving the video backgrounds the most, and the third environment preserving the video backgrounds the least.}
\end{figure*}

\begin{figure*}[h!]
    \centering
    \includegraphics[width=\linewidth]{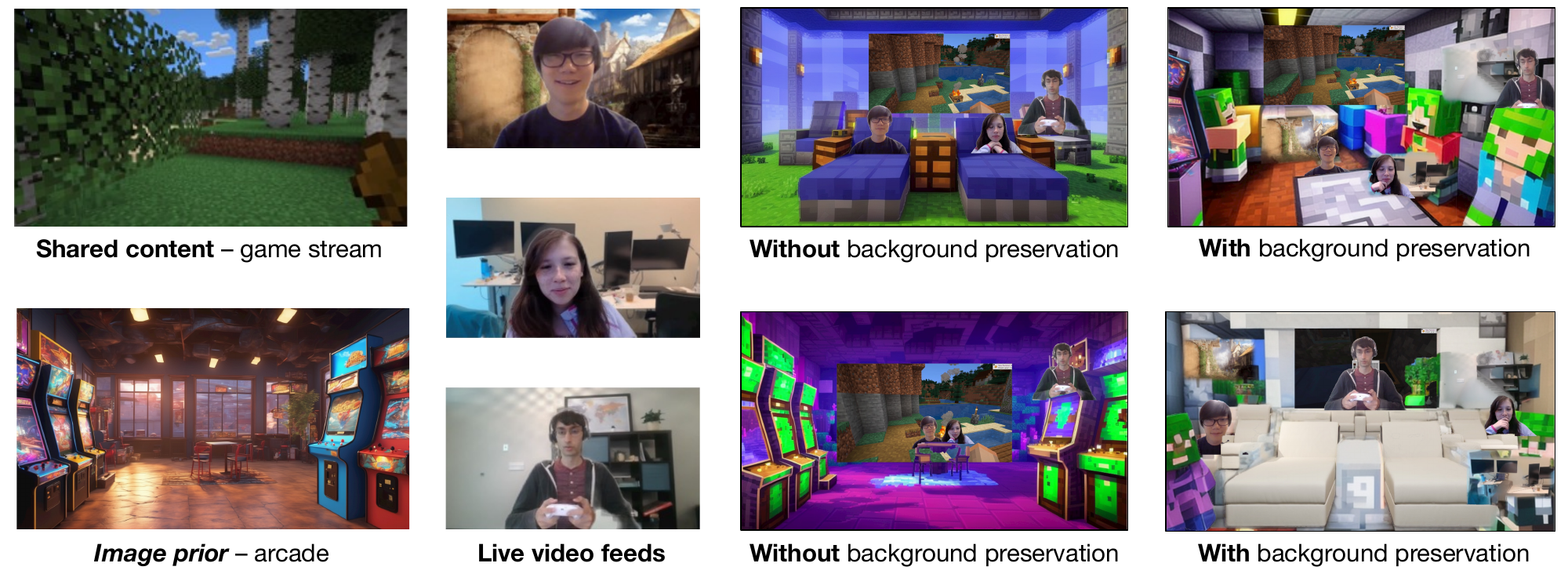}
    \caption{Game Stream Scenario: A gamer incorporates his Minecraft stream into a blended environment to make his viewers feel more connected to the gameplay. Participants compared four versions of spaces generated via image-to-image techniques, using two different priors and varying levels of background preservation.} 
    \label{fig:task2-envs}
    \Description{Overview of the "Game Stream" Scenario: A gamer incorporates his Minecraft stream into a blended environment to make his viewers feel more connected to the gameplay. Participants compared four versions of spaces generated via image-to-image techniques, using two different priors and varying levels of background preservation. This figure shows the four generated environments, of which two preserve the video backgrounds and two do not. Each environment has a unique placement of users and the Minecraft stream.}
\end{figure*}

\end{document}